# Real space visualization of entangled excitonic states in charged molecular assemblies

Jiří Doležal[1,2*], Sofia Canola[1*], Prokop Hapala[1], Rodrigo C. de Campos Ferreira[1], Pablo Merino[3,4*], Martin Švec[1,5*]

[1]Institute of Physics, Czech Academy of Sciences; Cukrovarnická 10/112, CZ16200 Praha 6, Czech Republic

[2]Faculty of Mathematics and Physics, Charles University; Ke Karlovu 3, CZ12116 Praha 2, Czech Republic

[3]Instituto de Ciencia de Materiales de Madrid; CSIC, Sor Juana Inés de la Cruz 3, E28049 Madrid, Spain

[4]Instituto de Física Fundamental, CSIC; Serrano 121, E28006 Madrid, Spain

[5]Regional Centre of Advanced Technologies and Materials; Šlechtitelů 27, CZ78371 Olomouc, Czech Republic

## Abstract

Entanglement of excitons holds great promise for the future of quantum computing, which would use individual molecular dyes as building blocks of their circuitry. Even though entangled excitonic eigenstates emerging in coupled molecular assemblies can be detected by far-field spectroscopies, access to the individual modes in real space will bring the much needed insight into the photophysics of these fascinating quantum phenomena. Here we combine tip-enhanced spectromicroscopy with atomic force microscopy to inspect delocalized single-exciton states of charged molecular assemblies engineered from individual perylenetetracarboxylic dianhydride molecules. Hyperspectral mapping of the eigenstates and comparison with calculated many-body optical transitions reveals a second low-lying excited state of the anion monomers and its role in the exciton entanglement within the assemblies. We also demonstrate control over the coupling by switching the assembly charge states. Our results reveal the possibility of tailoring excitonic properties of organic dye aggregates for advanced functionalities and establish the methodology to address them individually at the nanoscale.

## Introduction

Single optically active molecules offer key advantages as quantum emitters due to their very small dimensions, well-defined optical transitions and photostability[1]. In addition, they can be electrically driven to transduce electric currents into optical signals[2]. Coupling between such molecular emitters leads to entangled excitonic states and finely depends on the exact arrangement of the emitters at the nanoscale.[3,4] The distance and mutual orientation, together with the local nanoscopic environment and the charge state determine their absorption and emission properties. The optical response of chromophore clusters in the visible- to near-infrared region can be addressed by tip-enhanced scanning probe spectromicroscopies.[5-10] Coupling among excitons and energy transfer mechanisms have been studied on neutral chromophores, yet no control over the exciton delocalization by imposing an external electric field or charge state has been achieved at the level of individual assemblies. Using perylenetetracarboxylic dianhydride (PTCDA) *anions* we create small clusters that manifest exciton delocalization, that involves the first two excited states of the molecule. We identify their eigenmodes by a combination of scanning tunneling microscopy-induced luminescence (STML), atomic force microscope (AFM), time-dependent density functional theory (TD-DFT) and photon map simulations and we show eigenmode switching by charge state control.

## Results and Discussion

Single PTCDA molecules adsorb flat on three monolayer (ML) - NaCl/Ag(111) centered over a $Cl^-$ ion and align with the principal NaCl lattice directions. Under these conditions, when sufficiently decoupled from the metal substrate, PTCDA spontaneously takes up an extra unpaired electron to its lowest unoccupied molecular orbital,[11-13] becoming a molecular anion radical with a total spin S = 1/2 ($D_0^-$)[14]. In the nanocavity polarized above a threshold bias voltage of -2.1 V, the chromophores emit predominantly in the infrared region[15], which is manifested by a sharp peak at 1.332 eV (Fig.1a, taken at -2.5 V), accompanied by red- and blue-shifted sidebands at 1.363 and 1.303 eV. We attribute the main spectral line to the decay from the first excited state ($D_1^-$) of the PTCDA anion and the sidebands to vibrational features[16]. We base this notion on: i) measurements by radio-frequency phase fluorometry of the dynamics of the excitation that find an effective radiative lifetime below 70 ps, unexpectedly short for a triplet state and shorter than the excited neutral and trion effective lifetimes on ZnPc in comparable conditions.[17-18] ii) TD-DFT calculations predicting a strong emission of $D_1^- \rightarrow D_0^-$ at 1.62 eV, close to the experimental value. iii) observation of single-exciton state delocalization in assemblies of molecules, driven by Coulombic coupling, compatible only with the interactions among the anions.

Hyperspectral electroluminescence maps measured on a grid of points at constant height of the tip above the chromophore can be correlated with the theoretically calculated excited states.[19,20] The photon map of the main emission line (Fig.1c) taken with a metallic tip shows high emission intensity at the carbonyl-terminated ends of the PTCDA and can be assigned to a decay from the first excited state to the ground state of the anion ($D_1^- \rightarrow D_0^-$, see Fig.1b). It bears a transition dipole moment along the long axis of PTCDA (hereby denoted as *longitudinal mode*, see Fig.1d). For generating a theoretical photon map, we use the computed corresponding transition density and simulate the coupling of the excitation to the optical field of the laterally scanning nanocavity. Our simulation procedure has the capability to simulate photon maps from any given transition density (see Supporting Information). The resulting map for anionic PTCDA (Fig.1e) is in good agreement with the experimental observation. Subtle feature differences between the experiment and theory can be attributed to laterally inhomogeneous probability of charge injection into the frontier orbital.[19,21-23]

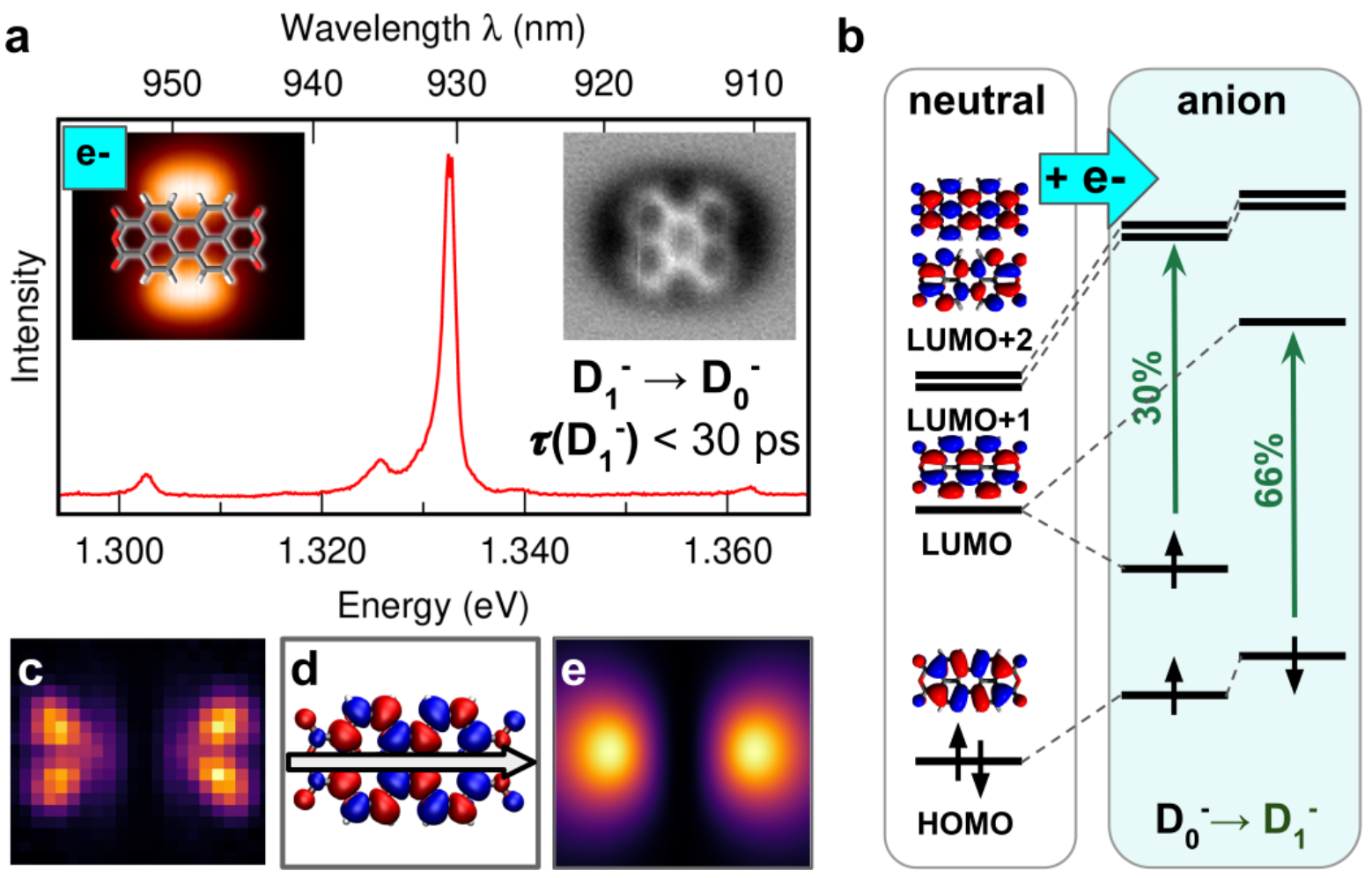

***Fig.1:*** *(a)* $D^-_1 \rightarrow D^-_0$ *spectral line and the lowest vibrational sidebands of the electroluminescence of PTCDA anion, taken at -2.5 V at the sample, 50 pA. The insets show tunneling current image at constant height, also taken at -2.5 V, with a PTCDA model overlay (left inset) and AFM constant-height image in the repulsive mode (right inset), showing the perylene backbone with atomic resolution. Both images are at the same scale (1.6 x 1.4 nm*$^2$*). (b) Energy level schemes with the corresponding orbitals from calculations on the PTCDA neutral and anion states. Green arrows are marking the main transitions involved in the first excited state* $D_1^-$ *and their weight in percent. (c) Experimental photon map integrated in the 1332±2 meV range, taken at -2.5 V and constant height and normalized. (d) Isosurface plots of the transition density of the PTCDA anion first excited state. The arrow denotes the transition dipole moment. (e) Simulated photon map based on the transition density in (d).*

We have assembled few-unit PTCDA anion clusters by thermally controlled diffusion.[11,12,24] The exciton delocalization in the assembly is manifested by a characteristic multiplicity of the exciton peaks (Davydov splitting) due to the emerging eigenmodes resulting from the interaction between the molecules. A non-linear parallel dimer configuration (Fig.2a-c) represents a geometry in which the longitudinal modes of PTCDA can couple efficiently. The spectra show a dominant contribution at 1.331 eV, with most intensity concentrated at the outer oxygen-terminations of the dimer, and a second less intense peak at 1.333 eV, which can be detected in the center of the dimer (Fig.2f). The doubly anionic dimer with one negative charge per molecule is the most stable charge configuration according to calculations (Supporting Information Table S1) and has a total spin S = 1 ($T_0^{2-}$). The first two excited states of the system, ($T_1^{2-}$ and $T_2^{2-}$), are dominated by excitations involving the same orbitals as the $D_0^- \rightarrow D_1^-$ monomer transition (Fig.2g) and thus can be understood as the in-phase and out-of-phase linear combinations of the longitudinal modes of the individual molecules (Fig.2e). The observation of the out-of-phase mode using the nanocavity, which is otherwise expected to be dark in the far-field measurement, is consistent with previous prediction[25]. The exciton states ordering, along with their energy spacing, provided by the calculations is confirmed by the very good agreement between the simulated and experimental photon maps (Fig.2f).

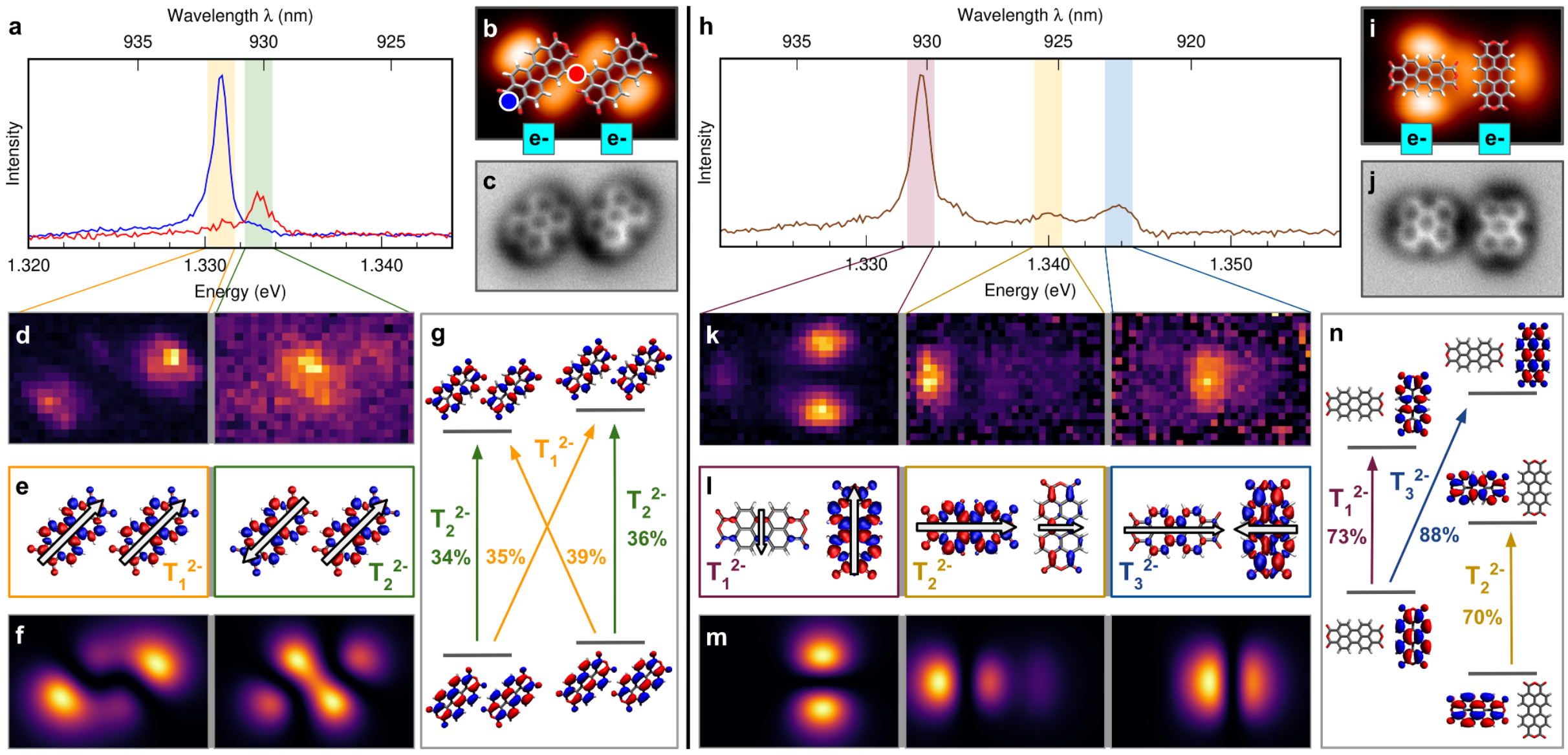

***Fig.2:*** *(a,h) Photon spectra of the parallel and perpendicular dimer emission, respectively. (b,i) Corresponding constant-height current images of the dimers at -2.5 V with model overlays. The charge state of each molecule is symbolized by a cyan square with e- within. The representative locations of the spectra in (a) for the parallel dimer are marked by their corresponding colors. (c,j) AFM constant-height images of the backbones taken at 5mV. (d,k) experimental photon maps integrated in the ranges marked in the spectra. (e,l) Isosurface plots of the transition densities of the lowest-energy excited states, assigned to the observed peaks in the spectra. White arrows denote schematically the direction of the longitudinal mode transition dipole moments of the individual molecules constituting the dimers. (f,m) Simulated photon maps based on the transition densities in e,l. (g,n) Scheme of the excitations with dominant contributions to the calculated lowest-energy eigenstates (weight in percent).*

We have investigated a second dimer motive that adopts a perpendicular arrangement of the molecules (Fig.2h-j). In the framework of the point-dipole approximation, the interaction between the longitudinal modes of the two perpendicular molecules is zero and therefore negligible coupling between exciton states of the two isolated molecules is expected. In contrast, the spectrum of the dimer shows multiple splitting of the main emission line, suggesting an additional exciton coupling mechanism. Three peaks at 1.333 eV (most intense), 1.340 and 1.344 eV are resolved, each generating a unique contrast in their corresponding photon map. Calculations of the doubly charged dimer, which is also the most stable in this case, show that lowest exciton states $T_1^{2-}$ and $T_2^{2-}$ (Fig.2l) are dominated by a transition that shares the characteristics with the $D_1^-$ monomer state localized on a single molecular unit (see Fig.2n), but with a non-negligible contribution from a higher energy mode ($D_2^-$) localized on the adjacent molecule (see Supporting Information Table S1). Indeed, we experimentally observed an additional line from a higher mode in single PTCDA, a transition at ~1.493 eV, when inspecting the molecule on 3ML-NaCl/Au(111) that can be directly assigned to the calculated $D_2^- \rightarrow D_0^-$ emission (see Supporting Information Fig.S3). According to both experiment and calculations, the $D_2^-$ excitation is energetically close to the $D_1^-$ state and has a transition dipole moment oriented perpendicular to the longitudinal mode (thereby *transversal mode*). Consequently the perpendicular arrangement of PTCDA molecules favors the coupling between longitudinal and transversal modes of the neighboring monomers. The transversal mode becomes the dominant contribution in the higher-energy state $T_3^{2-}$.

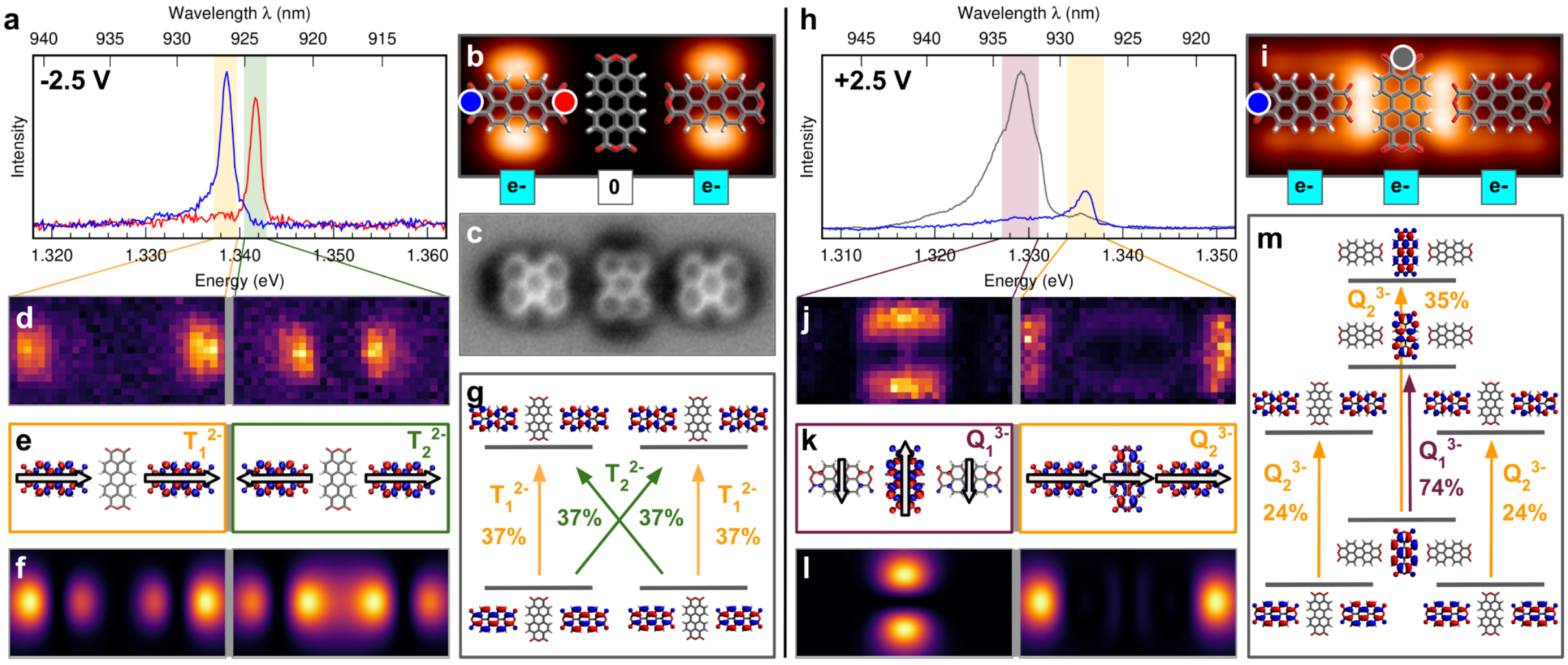

***Fig.3***: *(a,h) Photon spectra of the trimer taken at opposite polarities, -2.5 and +2.5 V respectively, representative of spectra in the locations marked by corresponding colors in (b,i), the constant-height current images of the dimers at -2.5 V with model overlays. The charge state of each molecule is symbolized by a cyan square with a value within. (c) AFM constant-height image of the backbones taken at 5mV. (d,j) the experimental photon maps integrated in the ranges marked in the spectra. (e,k) Isosurface plots of the transition densities of the lowest-energy excited states, assigned to the observed peaks in the spectra. White arrows denote schematically the directions of the transition dipole moments of the individual molecules constituting the trimers. (f,l) Simulated photon maps based on the transition densities in e,k. (g,m) Scheme of the excitations with dominant contributions to the calculated lowest-energy eigenstates (with weights in percent).*

We further target the role of the transversal mode by analyzing longer perpendicular arrangements. The PTCDA trimer (Fig.3) represents a simple extension of the perpendicular dimer. Surprisingly, the tunneling current image taken at negative bias (occupied states, Fig.3b) starkly contrasts with previous measurements on the monomer and dimers. The characteristic two-lobe pattern of the highest occupied level of a single molecule, when tunneling at -2.5 V, is absent from the central molecule. This suggests that the central molecule has a different charge state (neutral) than the peripheral ones (anions). Photon emission measurements show two well-separated emission lines at 1.338 and 1.342 eV indicating that only two longitudinal modes interact (similarly as in the parallel dimer).

Photon maps of the lines yield intensity patterns with maxima located at the outer and inner oxygen-terminations of the peripheral molecules, respectively. Computed excitation transition densities for a doubly charged trimer with $1e^-$ localized on each peripheral molecule - the lowest energy configuration of the trimer in DFT calculations - yield $T_1^{2-}$ and $T_2^{2-}$ eigenmodes as linear in- and out-of-phase combinations of the longitudinal modes of the charged molecules. The central neutral molecule is effectively excluded from the excitonic interactions in the trimer and acts merely as a spacer. Our calculations indeed confirm that in the neutral molecule there is no available transversal mode to couple with the longitudinal modes of the PTCDA anions.

Reversing the inactive role of the neutral molecule in the trimer can be achieved by controlling the electric field in the picocavity. By applying positive bias voltages we drive the system to a total charge of $3e^-$. The current image taken at +2.5 V shows a strong contribution to the tunneling current also at the central molecule. Photon spectra at positive polarities display a strong slightly broadened emission band at 1.329 eV (FWHM: 3.3 meV). Comparison of the experimental and theoretical photon maps, computed considering the trimer in a reference quartet spin state $Q_0^{-3}$ (see Supporting Information Table S2) unambiguously assigns the 1.329 eV line to the lowest excited eigenstate $Q_1^{-3}$. This excitation originates from the longitudinal mode of the central PTCDA weakly mixing with the transversal modes of its neighbors. The second excited mode $Q_2^{-3}$ is similar in energy and character to the first excitation of the doubly charged trimer $T_1^{-2}$, but in addition it incorporates coupling to the transversal mode of the central molecule. The bimodal charge state and the electrofluorochromic character of the PTCDA trimer can be further explained by DFT calculations. The $2e^-$ and $3e^-$ charge states have similar total energies, with the $2e^-$ being the lowest one in the gas phase (see Supporting Information Table S2). Positive strong bias voltages force the assembly in a triply charged ground state which leads to involvement of the central PTCDA in the exciton interaction. This demonstrates *reprogramming* of the excitonic system is possible by electric field control.

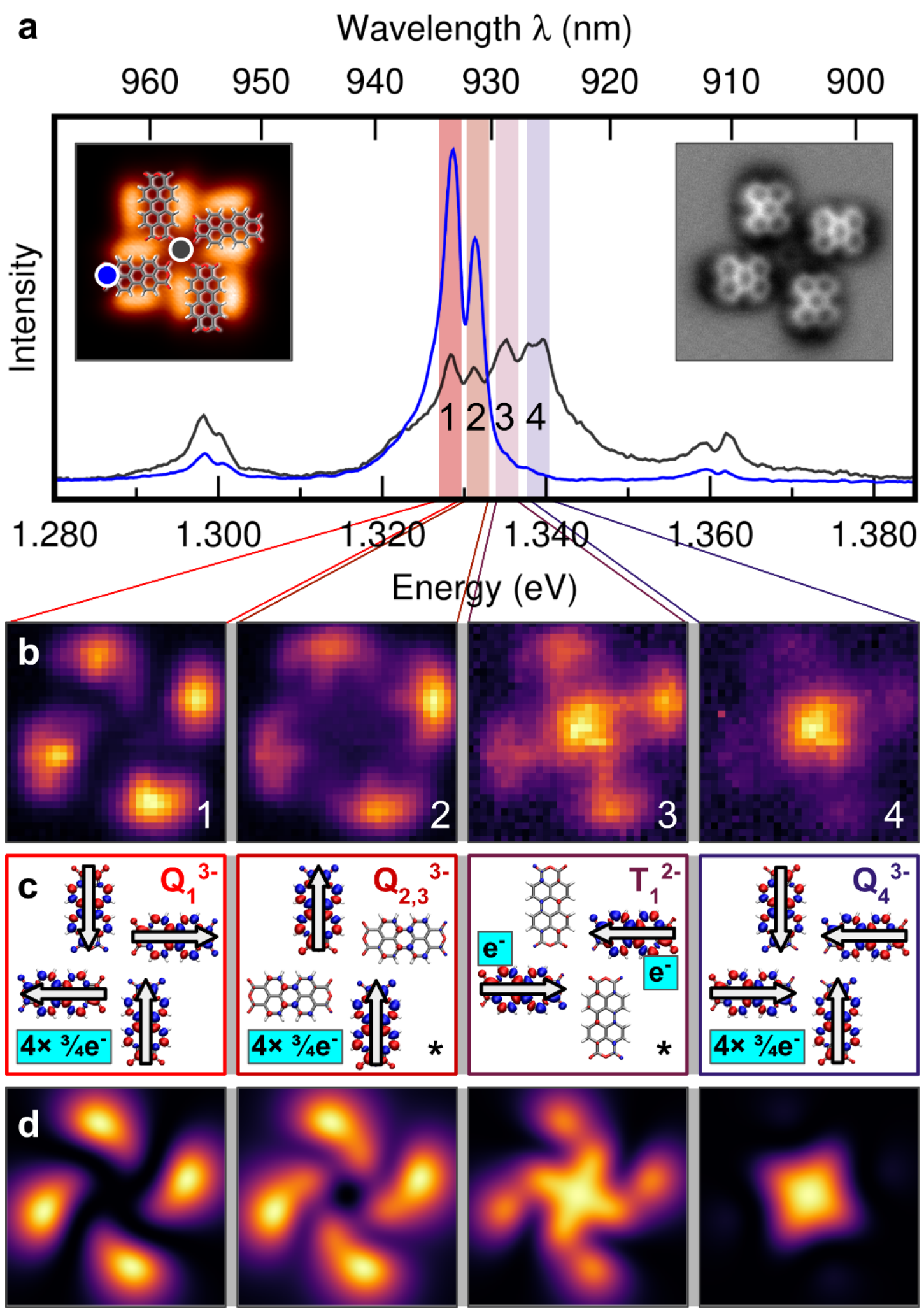

*Fig.4: (a) Photon spectra of the tetramer representative of the peripheral and center locations, marked in the current image taken at constant height (left inset). The bias voltage for the spectra and the current image is +2.5 V. The atomic model is overlaid. AFM high-resolution constant-height image (right inset) is taken at 5 mV (b) the experimental constant-height photon maps integrated in the ranges marked in the spectra as 1-4. (c) Isosurface plots of the transition densities of the lowest-energy excited states of the 3e$^-$ and 2e$^-$ configurations, assigned based on the observed photon maps. White arrows denote schematically the direction of the transition dipole moments of the individual molecules. (d) Simulated photon maps based on the transition densities in c. The maps 2 and 4 are obtained by averaging the corresponding two 90° mutually rotated photon maps.*

To further explore the exciton delocalization and the relation to the charge state we have assembled larger PTCDA aggregates. In the highly symmetrical tetramer introduced in Fig.4 the total number of extra electrons appears to reach a local equilibrium where the Coulomb repulsion between charges on adjacent molecules counteract the propensity of PTCDA to take up an electron. In large assemblies integer charges do not necessarily need to be localized at particular molecules and instead can reside in delocalized orbitals of the whole ensemble. The calculation of the tetramer reveals a near degeneracy between the double (2e$^-$) and triple (3e$^-$) negatively charged configurations. (see Supporting Information Table S2). For the $2e^-$ configuration, the charge is localized on two opposite molecules and for the $3e^-$ case, it evenly distributes among the four molecules ($3/4e^-$ each). In the photon +2.5V spectra we resolve four modes at energies, 1.328, 1.332, 1.335 and 1.339 eV (denoted as 1-4, respectively), which can be theoretically explained by considering the system alternating between the $3e^-$ and $2e^-$ states. The experimental photon maps and the relative energy order of modes 1, 2 and 4 can be rationalized by quartet ($Q_n^{-3}$, n = 1,2,3) excited states of the $3e^-$ configuration. In the highest energy mode (4, $Q_4^{-3}$) the longitudinal modes of the monomers couple with individual dipoles pointing inwards, producing a photon map with a characteristic intensity maximum in the center. Conversely, the two lowest energy modes (1 and 2, $Q_1^{-3}$ and $Q_2^{-3}$) consist of alternative orientations of the longitudinal transition dipole moments which give photon maps with four lobes located in the outer part of the tetramer. Mode 3 showing a spatial distribution with high intensity on both the periphery and the center of the tetramer is well described by the lowest energy triplet excitation of the doubly charged assembly, $T_1^{-2}$. For this case as well as for the $Q_2^{-3}$, the two rotationally equivalent configurations are superimposed in order to simulate the degeneracy of the $2e^-$ localization. We believe that the tetramer, constantly disturbed from the equilibrium by injection of charges, oscillates between the two states and therefore both contribute to the observed photon maps.

**Conclusion**

In summary, based on a combined experimental and theoretical approach, we have created elementary entangled single-exciton states from the PTCDA anions and identified their eigenmodes, including states otherwise considered dark. We have revealed the role of the first two excited states $D_1^-$ and $D_2^-$ of the anion in the dipolar coupling between monomers and demonstrated control over the excitonic properties of the assemblies by atomically precise nanopositioning and electrical charge control.

**Associated content**

**Supporting Information**: sample preparation and STML/AFM measurements; DFT and TD-DFT calculations; simulation of the photon maps; RF-PS measurements; Assembly formation on NaCl/Ag(111); single PTCDA on 3 layers of NaCl/Au(111) at positive bias voltage; determination of the adsorption geometries with AFM; normalization of the spectra and the photon maps; calculations on single molecule and aggregates.

**Author Information**

Corresponding Authors:

*E-mail: M. Švec (svec@fzu.cz), J. Doležal (dolezalj@fzu.cz), S. Canola (canola@fzu.cz), P. Merino (pablo.merino@csic.es).

**Acknowledgment**

SC, RCCF, MŠ and JD acknowledge the Czech grant agency funding no. 20-18741S and the Charles University Grant Agency project no. 910120. PM thanks the ERC Synergy Program (grant no. ERC-2013-SYG-610256, Nanocosmos) for financial support and the "Comunidad de Madrid" for its support to the FotoArt-CM Project (S2018/NMT-4367) through the Program of R&D activities between research groups in Technologies 2013, co-financed by European Structural Funds.

# Supporting Information for

## Real space visualization of entangled excitonic states in charged molecular assemblies

Jiří Doležal[1,2*], Sofia Canola[1*], Prokop Hapala[1], Rodrigo C. de Campos Ferreira[1], Pablo Merino[3,4*], Martin Švec[1,5*]

[1]Institute of Physics, Czech Academy of Sciences; Cukrovarnická 10/112, CZ16200 Praha 6, Czech Republic

[2]Faculty of Mathematics and Physics, Charles University; Ke Karlovu 3, CZ12116 Praha 2, Czech Republic

[3]Instituto de Ciencia de Materiales de Madrid; CSIC, Sor Juana Inés de la Cruz 3, E28049 Madrid, Spain

[4]Instituto de Física Fundamental, CSIC; Serrano 121, E28006 Madrid, Spain

[5]Regional Centre of Advanced Technologies and Materials; Šlechtitelů 27, CZ78371 Olomouc, Czech Republic

Sample preparation and STML/AFM measurements.

All measurements were performed in an ultrahigh vacuum system with base pressure below $5\times10^{-10}$ mbar. The Ag(111) and Au(111) clean surfaces were prepared by standard procedure - cycles of 0.5-1.0 keV $Ar^+$ ion sputtering and annealing to 550°C. NaCl was evaporated from a source at 607°C on the surface kept at 120°C during 3-5 min. The duration of deposition was tuned in order to obtain NaCl islands with 2-4 atomic layers. Subsequently the sample was inserted into the microscope head attached to a liquid He bath cryostat. For the entire study we used an AFM sensor with a base resonant frequency of ~30 kHz and a PtIr tip, sharpened by a focused ion beam before inserting into the scanner. Tips were cleaned and coated by Ag or Au material by applying voltage pulses and controlled nanoscopic indentations into the clean substrate. The AFM images were taken in frequency modulation mode at constant amplitude of 50 pA with CO-functionalized tips. In the optical spectroscopy measurements, the photons were collected by a lens in the vicinity of the tunneling junction and directed out of the cryostat through a set of viewports. The outgoing beam was refocused into an optical fiber bunch leading to Andor Kymera 328i spectrograph equipped with a CCD sensor (Andor Newton DU920P-BEX2-DD). Photon spectroscopy measurements were taken with gratings of 300, 600 and 1200 grooves/mm, which at 1.33 eV provide spectral resolutions better than 3.6, 0.8 and 0.5 meV, respectively.

DFT and TD-DFT calculations.

All the quantum mechanical calculations were performed with the Gaussian16 package.[S1] The ground state calculations were done with density functional method (DFT) employing wB97XD functional[S2] and 6-31G* basis set. The presence of a more stable open shell solution was checked via the *stable* option and for open-shell systems UDFT has been used. The singlet excited state of neutral systems were computed with time-dependent (TD) DFT at the TD-wB97XD/6-31G* level of theory. For the triplet neutral and all the charged states, the Tamm-Dancoff approximation (TDA) has been used (TDA-ωB97X-D/6-31G*). The triplet to singlet excited states transitions were computed with the *50-50* option. PTCDA single molecule was optimized *in vacuo* in its neutral and anion ground states. The vertical excited states were assessed at the ground state equilibrium geometry and the emission properties were obtained by optimization of the corresponding excited state geometry (Table S1, Fig.S7). All aggregate model structures were built on the basis of the information provided by the analysis of the AFM images (Fig.S6), using the optimized PTCDA neutral geometry as basic units and neglecting the role of the substrate. For each aggregate several anionic states have been considered, by varying the number of total negative charges, and the most stable have been selected for further calculations (Table S2). In general, for each given total charge state, all relevant spin multiplicities have been computed with UDFT and provided near degenerate energies, due to a negligible spin interaction between the unpaired electrons. The total amount of charge localized on each molecular unit has been estimated based on Mulliken population analysis, by summing the partial charges on each atom within the molecule.[S3] The aggregate exciton states (Tables S3-S8) were computed as vertical excitations, at a fixed geometry, on a selected spin state for each charge arrangement listed in Table S2 along with their oscillator strengths *f*, which characterize the linear absorption response of the system in the far field. The transition densities of selected states are used to simulate the corresponding photon maps. The transition densis were generated employing Multiwfn software (grid spacing 0.2 Bohr; plot with isosurface value set to 0.0001 e/Bohr$^3$).[S4]

Simulation of the photon maps

To compare the experimentally measured photon maps with the corresponding theoretical predictions we developed a computationally efficient and universal module within the freely available Probe-Particle package[S5], which calculates the interaction between the electric field of a tip-enhanced plasmon, and a molecular or aggregate exciton in the sample. Following the approach developed by Neuman *et al.,*(based on Eq.(1) in ref. S6), we evaluate the coulomb integral *g(r)* between the transition density of the excitons $\rho_S$ of the sample (e.g. the molecular cluster) and the electric field $V_{TIP}$ of the metallic tip plasmon:

$$g(r) = \int_{r'} \rho_S(r') V_{TIP}(r - r')\, dr' \qquad (1)$$

where *r* represents the relative position of the tip over the sample, corresponding to a pixel in the simulated photon map and *r'* is a grid coordinate.

We use the calculated function $|g(r)|^2$ directly to generate the photon maps. This represents a simplification of Eq.(1) from ref. S6 by removing its energy dependence, assuming that within the narrow energy range of molecular exciton the spectrum of the plasmon is approximately constant. This approach is sufficient to study the spatial dependence of the coupling (i.e. the dependence on tip position *r*) and for the purpose of comparison with the normalized experimental photon maps in which the background and spectral dependence is largely suppressed. In our simulation, $V_{TIP}$ is currently approximated by damped multipole expansion:

$$V_{TIP}(r) = \sum_k C_k ((z - z_0)/|r|)^{(k-1)} (1/(|r|^2 - R^2)^{(k/2)}) \qquad (2)$$

where $k$ controls the order of multipole ($k$ = 1 monopole, $k$ = 2 dipole, etc.), $z_0$ represents the distance of the tip (the centre of the $V_{TIP}$ multipole) from the sample, and $R$ denotes the finite width of the tip (i.e. Lorentzian decay for the dipole). In line with the approach employed in ref. S6, we approximated the tip field plasmon function $V_{TIP}$ as an electric dipole oriented in the *z*-direction with $z_0 = R = 5$ Å. However, we verified that photon maps are qualitatively similar if simulated using an electric monopole or a dipole oriented along the *z*-axis.

The transition densities $\rho_S$ of the aggregates were obtained directly from the quantum-chemical calculations (see below for details). Both $V_{TIP}$ and $\rho_S$ are sampled on a regular real-space grid with resolution ~0.2 Å before the simulation, which allows us to easily update our model in future to include e.g. more a realistic shape of the tip.

To calculate the convolution in Eq.(1) efficiently, we employ fast fourier transform (FFT). This makes our simulation program very efficient and general. For arbitrary spatial distribution of $V_{TIP}$ stored on a real space grid it can simulate photon-maps in a few seconds on a standard PC.

RF-PS measurements

For the RF-PS spectroscopy on a PTCDA anion we used the methodology described in our previous work.[S7] The transmission of the wiring was calibrated at 200 MHz using broadening of a plasmon high-energy cutoff at 1.8 V on a clean substrate.[S8] The time frame is 50 ns and we used 64 ps bin width. For filtering of the emission line we used a hard-coated 25 nm bandpass filter (Edmund optics) with center wavelength 925 nm. Histograms are accumulated for approximately 20 min, depending on the strength of the modulation and the resulting signal-to-noise ratio in the waves (Fig.S1). At the bias of -2.1 V (onset of the luminescence) the obtained effective lifetime of the excited anionic state is within the error of the measurement which is <70 ps in this case.

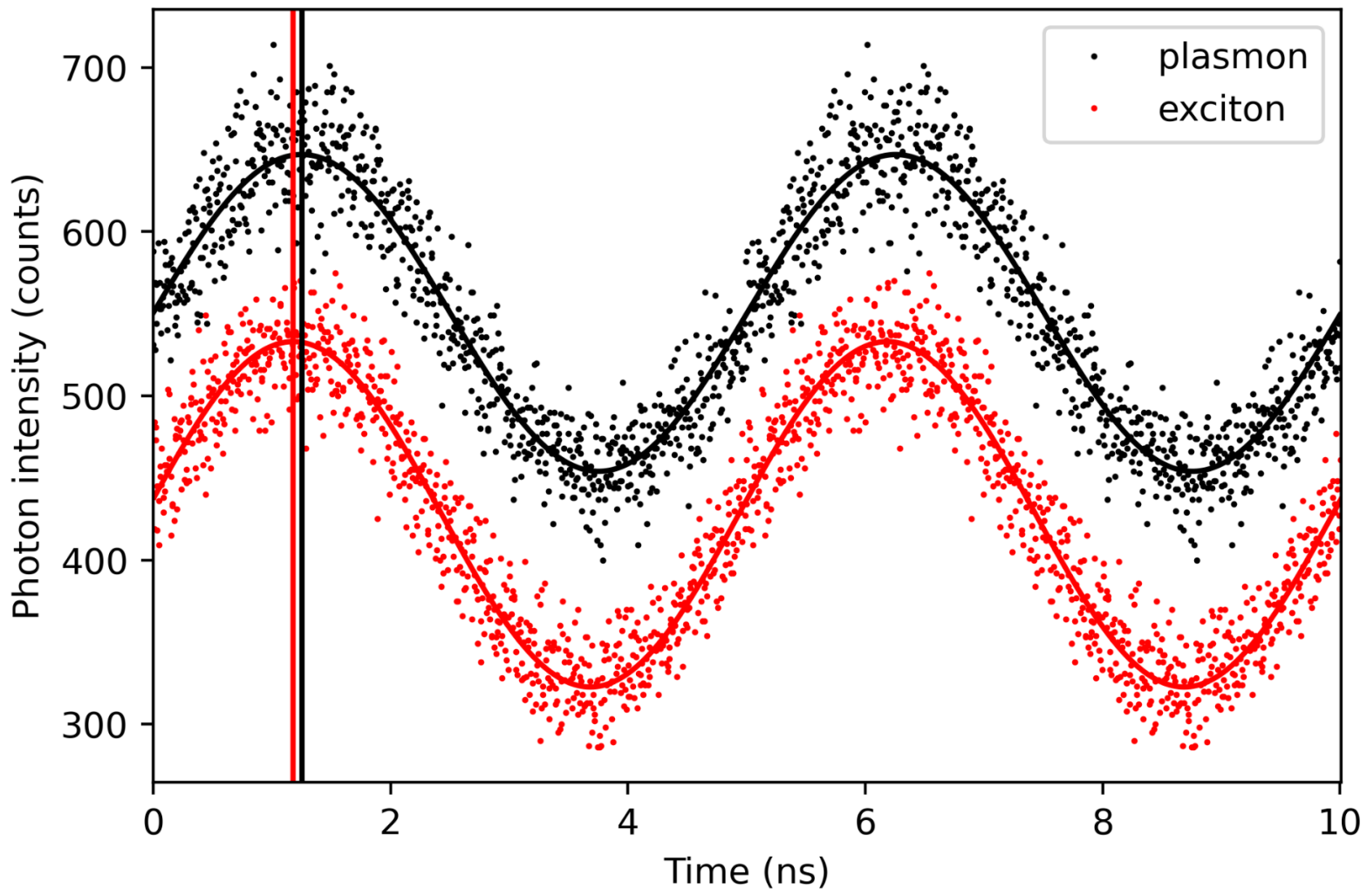

**Figure S1.** Radio-frequency phase-shifted waves obtained for a plasmonic reference (black) $V_{DC}$ = 1.5 V, $I_t$ = 1 nA, $V_{AC}$ = 200 mV, integration time 1200 s on Ag(111). First excited anionic state (red) was measured with $V_{DC}$ = -2.1 V, $I_t$ = 50 pA, $V_{AC}$ = 200 mV, integration time 1200 s at oxygen termination position on a single PTCDA on 3ML-NaCl. Sinus function with calculated phase-shift, amplitude, constant term and period as parameters is plotted with a solid line for both plasmonic reference and excited state. The difference between the reference and the excited state is marked by a vertical line.

Assembly formation on NaCl/Ag(111)

To generate the molecular assemblies of different configurations we have tested several methodologies. While molecular manipulation with the tip of the AFM/STM could be used to form molecular dimers, in order to efficiently form larger assemblies, we have adopted the thermally activated process previously described in the literature.[S9-S11] We successfully applied this concept to the PTCDA deposited on NaCl on Ag(111) (evaporation temperature was 380°C, substrate at 10 K) and subsequently annealed for 1 min by taking the sample with a manipulator (at RT) and putting it in contact with the LN cryostat shield. With the sample cooled back below 10 K, the STM topography shows an abundance of small 2D molecular assemblies (examples at low coverage can be seen in Fig.S2). Here the driving interaction of the self-assembly stems from the attraction between the carbonyl oxygens and the hydrogen-termination along the sides of the perylene backbone of the molecules, which pushes the nearest neighbors in the clusters to perpendicular arrangements. The result of the process is sensitive to the concentration of PTCDA on the surface, annealing duration and the number of NaCl layers on which the molecules diffuse. Higher concentrations and higher annealing times tend to produce larger islands (we have prepared clusters of up to 20 units); the largest islands tend to be formed at 3 layers of NaCl compared to 2 and 4 layer NaCl.

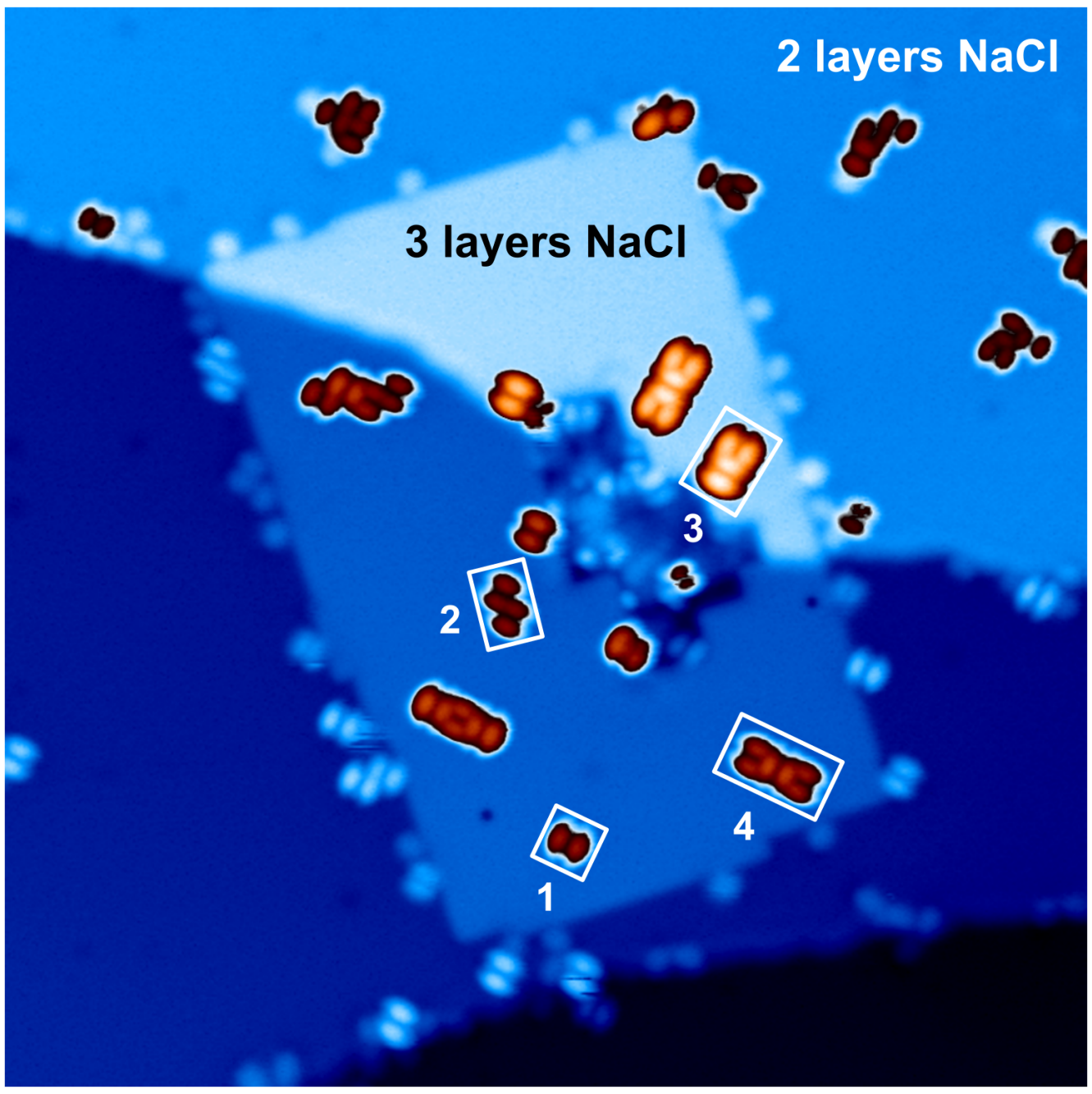

**Figure S2.** Example STM topography overview of the PTCDA aggregates on 2ML- and 3ML-NaCl/Ag(111) system, formed by allowing diffusion of the molecules. White rectangles mark a monomer (1), two types of dimers (2,3) and a trimer (4) that were the subject of this study. The image parameters were 40 x 40 nm$^2$, 1.2 V.

<u>Single PTCDA on 3 layers of NaCl/Au(111) at positive bias voltage</u>

For the lack of features and a low signal yield in the photon spectra of PTCDA at energies higher than the anion first excited state, we performed an additional measurement at positive bias voltages. However, under these conditions the single molecule experiences instability. Nevertheless, we have discovered that on NaCl/Au(111) it remains relatively stable and therefore we were able to perform the photon spectroscopy and hyperspectral mapping at a single molecule on this surface (Fig.S3). We found the dominant contribution of the first excited anion state at 1.332 eV and a number of higher-energy peaks with large plasmonic background. In particular, the spectra taken over the H-terminated sides present a significantly larger contribution of the signal at 1.493 eV, with respect to the intensity of the first excited state. A photon map at this energy (Fig.S3c reveals a distribution of the intensity in the real space, indicative of an excited state independent of the $D_1 \rightarrow D_0$ transition. We link this feature with the decay of a second excited state ($D_2$), based on the comparison with the theoretical prediction (Fig.S3d,e), which associates it with a transition dipole moment perpendicular to the longitudinal plane of the molecule, resulting in a characteristic photon map with the intensity localized above the H-terminations of the molecule.

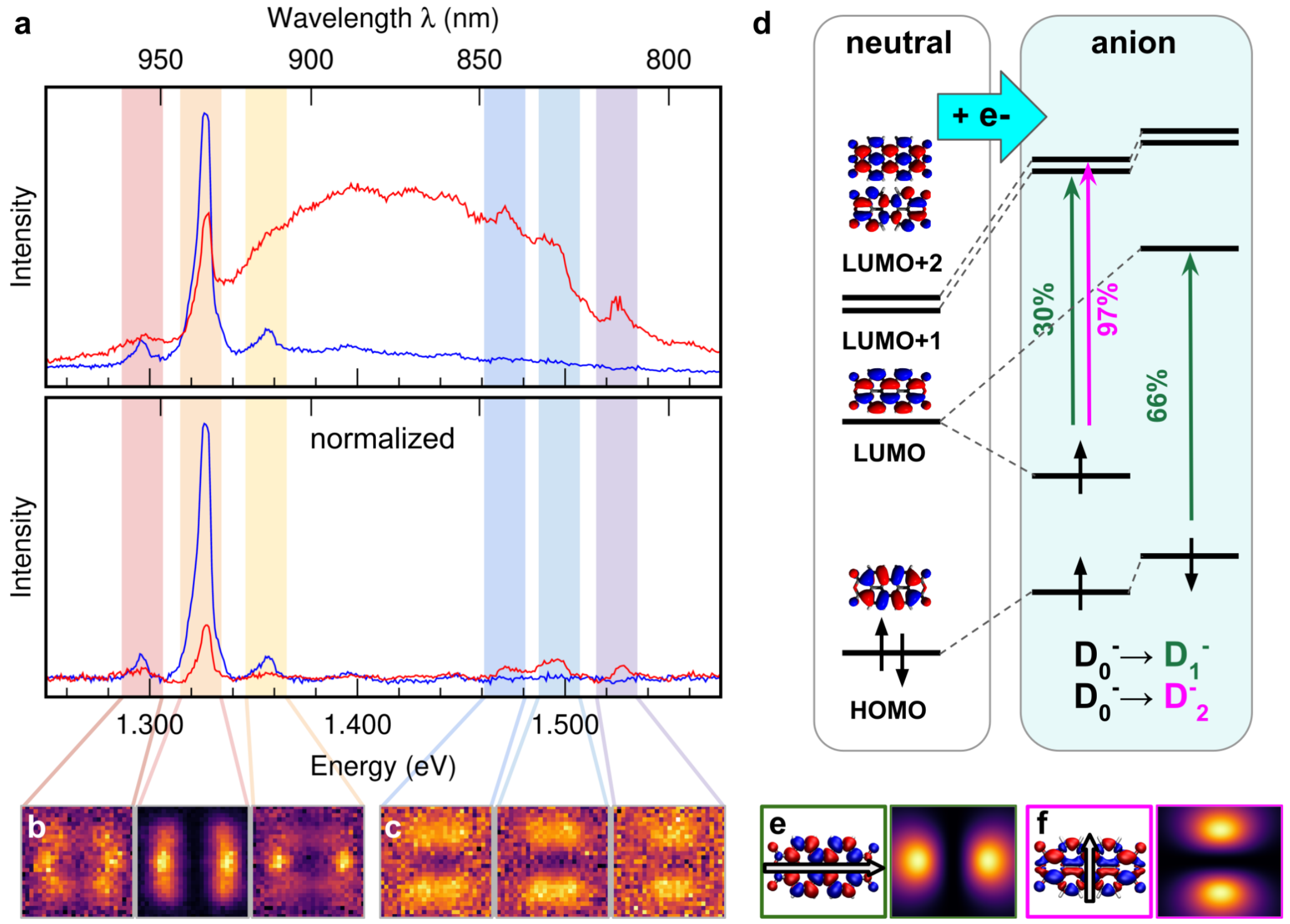

**Figure S3.** (a) Averaged electroluminescence spectra of PTCDA anion measured on 3ML-NaCl/Au(111), before and after (top and bottom panels, respectively) normalization by the plasmon. Measurement parameters were set to +2.5V, 100pA. (b,c) photon maps evaluated in the spectral ranges denoted in (a). (d) Energy level schemes and corresponding orbitals obtained from calculations on the PTCDA neutral and anion states (wB97XD/6-31G*). Green and magenta arrows are marking the main transitions involved in the first two anion excited states $D_1^-$ and $D_2^-$ and their weight in percent. (e,f) transition densities and their corresponding simulated photon maps of the $D_1^-$ and $D_2^-$ states, respectively.

Determination of the adsorption geometries with AFM

For a single PTCDA molecule, both types of dimers and the tetramer, we performed a geometrical registration with the substrate (Fig.S4). We used CO-functionalized tips to scan the molecule and the surrounding areas of NaCl substrate, each at relative tip-sample distances that yield the atomic resolution. By extrapolating the lattice of the NaCl, we determined the preferred positioning of the PTCDA, which is always on top of a $Cl^-$ ion, with the two principal mirror planes aligned with the Cl rows.

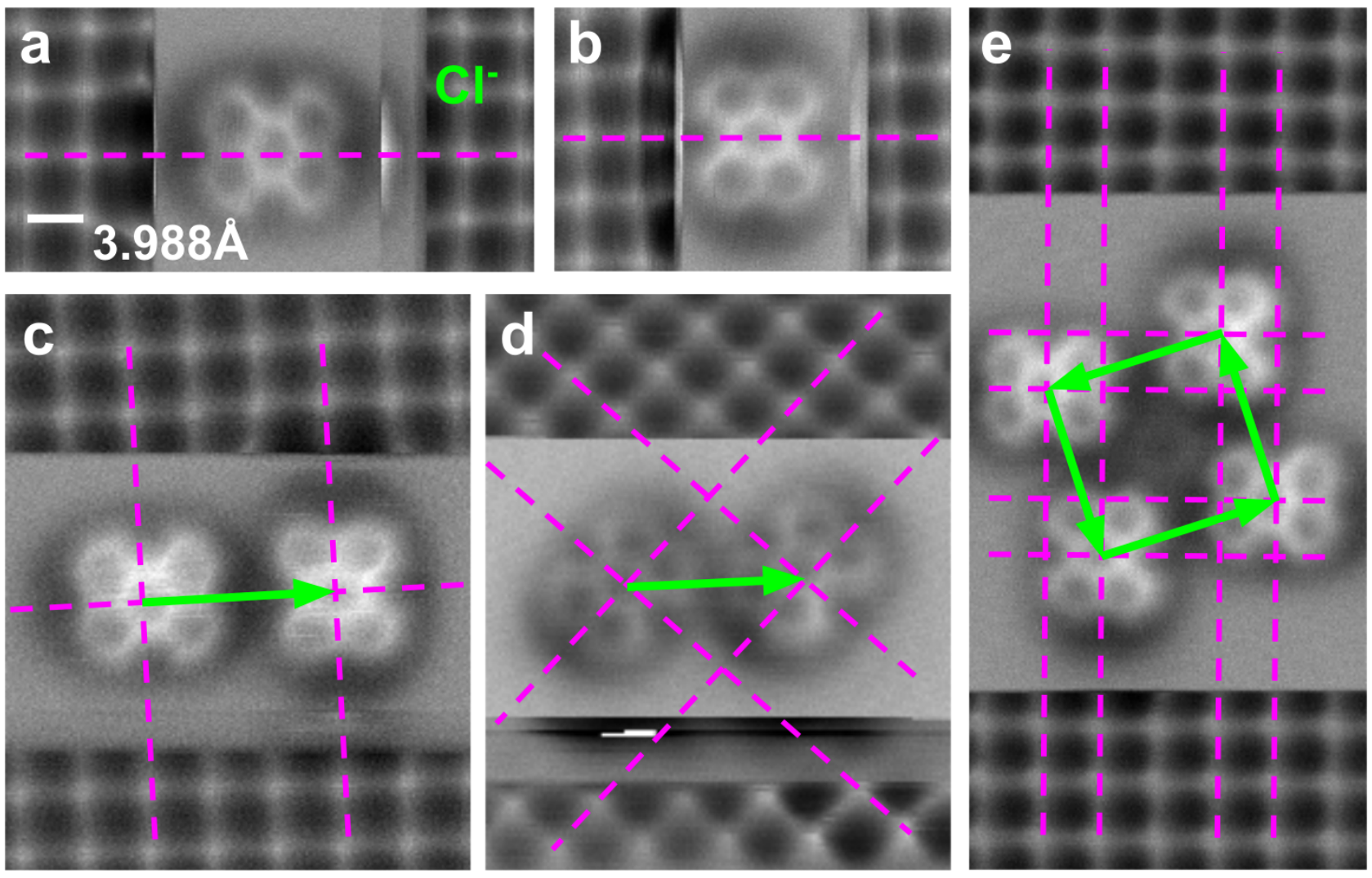

**Figure S4.** AFM analysis of adsorption configurations of the single PTCDA molecules (a), perpendicular dimer (b), parallel dimer (c) and tetramer (d) on NaCl(001) layers on Ag(111). Purple dashed lines mark the Cl rows which cross the centers of the PTCDA, green vectors denote the mutual displacements of the molecules in the aggregates. The bright spots in the lattice correspond to $Cl^-$ ion locations. All images have been taken with CO-functionalized tips at bias voltage of 5mV. The absolute scale of all images can be unambiguously derived from the visible surface lattice of NaCl(001).

Normalization of the spectra and the photon maps

The individual electroluminescence spectra taken above the molecules are shaped by the spectral character of the nanocavity and the tunneling current, which are inherently dependent on the precise position of the tip relative to the sample with the molecule, in addition to the instrumental function of the optical detection line. This will affect the intensity of emission lines as well as the background in any STML spectrum and consequently the contrast of the photon maps. Experimentally it was observed that the influence on the background intensity is partly mitigated on 3ML of NaCl in contrast to 2ML (as shown in our previous work[S12]), likely due to a weaker direct tunneling of electrons between the tip and the metal below the NaCl. However in the measurements in this study, performed on 3ML of NaCl the background still plays a non-negligible role, which prevents correct interpretation of weaker excitonic contributions. Therefore we seek a robust normalization procedure that would reliably correct the spurious factors modulating a hyperspectral map.

We assume that the detected photons originate either from plasmons (excited by inelastic electrons directly transported across the tip-molecule-NaCl-metal system), or from the excitons (generated by charge injection into the molecule), as depicted in the scheme in the inset of the Fig.S5. Generation of such plasmonic and excitonic photons would be in a first approximation proportional to the intensity of the flow of charges (current) and the effectiveness of coupling to the nanocavity according to the actual tip-sample geometry. Detected spectra of the excitons and plasmons will finally be modulated by the spectral efficiency of the optical detection setup. We represent the net effect of all these factors by a general modulation function $\Phi(r, E)$, which depends on the tip position $r$ and energy $E$. The detected intensity of the photons can be then written as

$$I'(r, E) = \alpha \cdot I(r, E) \cdot \Phi(r, E) + \beta \cdot p(E) \cdot \Phi(r, E) \quad (3)$$

where the $I(\boldsymbol{r},E)$ is the excitonic emission spectrum, $p(E)$ the plasmonic response of the nanocavity material and α and β the respective exciton and plasmon yields. For simplicity we approximate $p(E)$ as a constant and substitute $\beta \cdot \Phi(r, E)$ for $\Phi'(r, E)$, which represents the effective measured plasmonic background. With this we obtain

$$I(r, E) \propto [I'(r, E) - \Phi'(r, E)]/\Phi'(r, E) \quad (4)$$

By estimating the $\Phi'(r, E)$ for each position $\boldsymbol{r}$ in the hyperspectral map, we can recover the excitonic spectra and normalize the photon maps for such $\boldsymbol{r}$ and $E$ that yield a nonzero $\Phi'$. For estimation of the plasmonic background, we exploit the continuous character of the plasmonic contribution as opposed to the excitonic signal, which consists of individual lines and can thus be relatively simply distinguished. An example of the photon map normalization is provided in the Fig.S6, which shows the perpendicular dimer before and after the process, including the intermediate step of the background subtraction. The comparison of the raw and processed photon maps to the theoretical prediction (Fig.2k) demonstrates the necessity of this step in suppressing the background signal of non-excitonic origin.

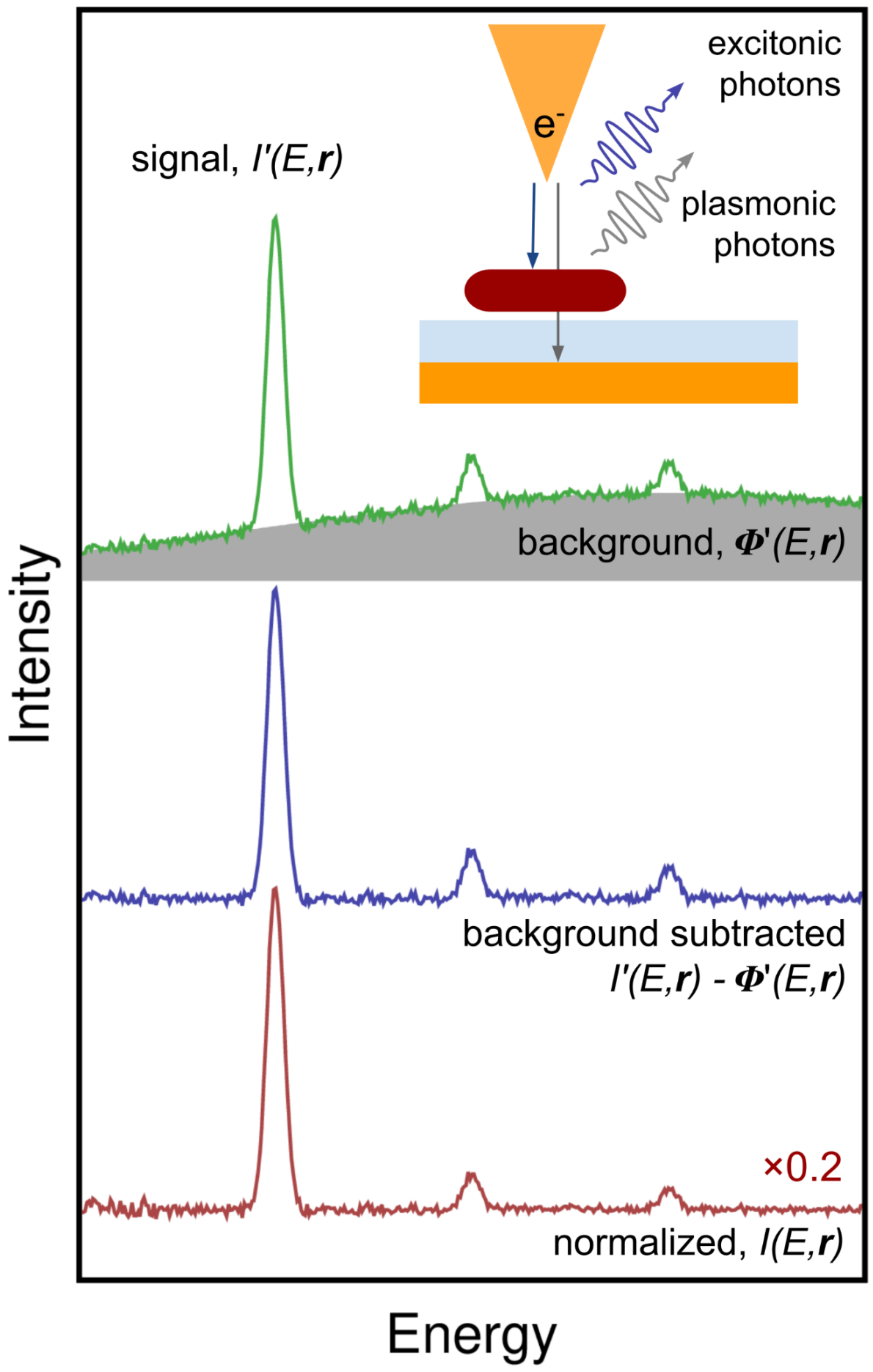

**Figure S5.** A visual explanation of the normalization procedure used for individual spectra comprising the hyperspectral maps. The detected signal $I'(r, E)$ is plotted by green color, the effective plasmonic background $\Phi(r, E)$ by grey filled area. The excitonic part of the signal, plotted in blue, is divided by this general background to obtain the final normalized excitonic spectrum $I(r, E)$ (in red). The inset schematically depicts the basic concept of the electrons tunnelled directly between the tip and the metal substrate, inducing plasmons, or being captured by the molecule, and leading to exciton formation and radiative decay.

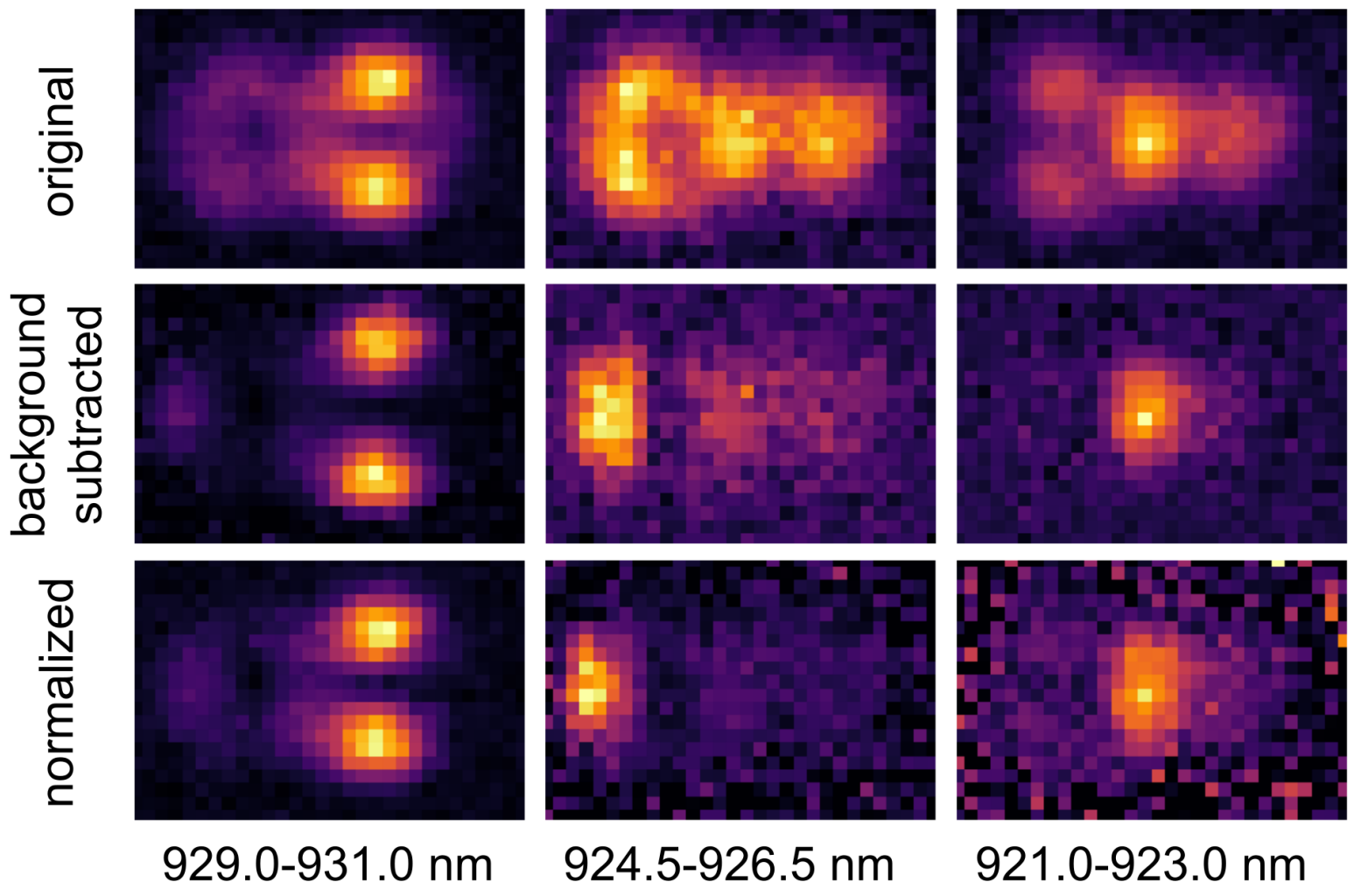

**Figure S6.** An example of the background subtraction and normalization of the hyperspectral maps taken at the perpendicular dimer.

Calculations on single molecule and aggregates

For the neutral molecule, the first excited state is represented by the HOMO→LUMO transition, with energy value 2.50 eV at the full TD-DFT level in excellent agreement with previously reported experimental data at 2.45 eV.[S13] The anion displays two low-lying excited states close in energy, with emission at 1.62 and 1.80 eV, respectively. These values are slightly overestimated compared to the experimental emission (at 1.332 eV and 1.493 eV), consistently with that of the neutral excited state predicted at 2.79 eV when employing the TDA approximation. The corresponding transition dipole moments are oriented along the long and short molecular axes, respectively (see Fig.1 and Fig.S3). The calculation locates the triplet to singlet emission to lower energies, around 1.18 eV, also in agreement with the previous work.[S13]

The aggregate exciton states computed with TD-DFT are linear combinations of single excitations involving orbitals delocalized on the whole system. In these PTCDA systems, the aggregate orbitals result from linear combinations of the molecular anionic orbitals localized on each molecular unit. Thus, for each aggregate orbital, by a simple visual inspection, the dominant contributions to the combinations of molecular anionic orbitals can be identified. On the basis of this simple analysis it is possible to relate the calculated aggregate excitonic states (along with their corresponding transition densities) to the excitations localized on each molecular unit (longitudinal and transversal modes), whose interaction generates the excitonic states. The lookup tables and corresponding orbital energy schemes are summarized in Tables S3-S8 and Figs.S8-S13. We denote the molecular units in the aggregates with letters (*a-d*, see Table S2). The notation of the aggregate orbitals is chosen to reflect the base monomer orbital (H, L, H+1, L+1, etc.), the sign of the orbital linear combination (sign present in the upper index), and localization/delocalization on particular molecular units (a-d in upper indices). The spin branches are marked as $\alpha$ and $\beta$.

| charge | transition | absorption E/eV (*f*) | absorption wavefunction[c] | emission E/eV (*f*) |
|---|---|---|---|---|
| neutral | $S_0 \leftrightarrow S_1$[a] | 2.91 (0.737) | 0.98 (H → L) | 2.50 (0.716) |
| | $S_0 \leftrightarrow T_1$[b] | 1.77 (0) | 0.90(H → L) | 1.18 (0) |
| anion[b] | $D^-_0 \leftrightarrow D^-_1$ | 1.84 (0.020) | 0.66(H→L) + 0.30 (L→ L+1) | 1.62 (0.162) |
| | $D^-_0 \leftrightarrow D^-_2$ | 2.08 (0.072) | 0.92 (L→L+2) | 1.80 (0.076) |

[a] Calculation with TD-wB97XD/6-31G*. [b] Calculation with TDA-wB97XD/6-31G*.
[c] Molecular orbital naming refers to the order in the neutral molecule (see Fig. 1 main text).

**Table S1.** Computed emission energies (E), oscillator strengths (*f*) and wavefunctions composition (wf), including the most relevant coefficients and orbitals involved in the dominant excitations) for the isolated PTCDA molecule.

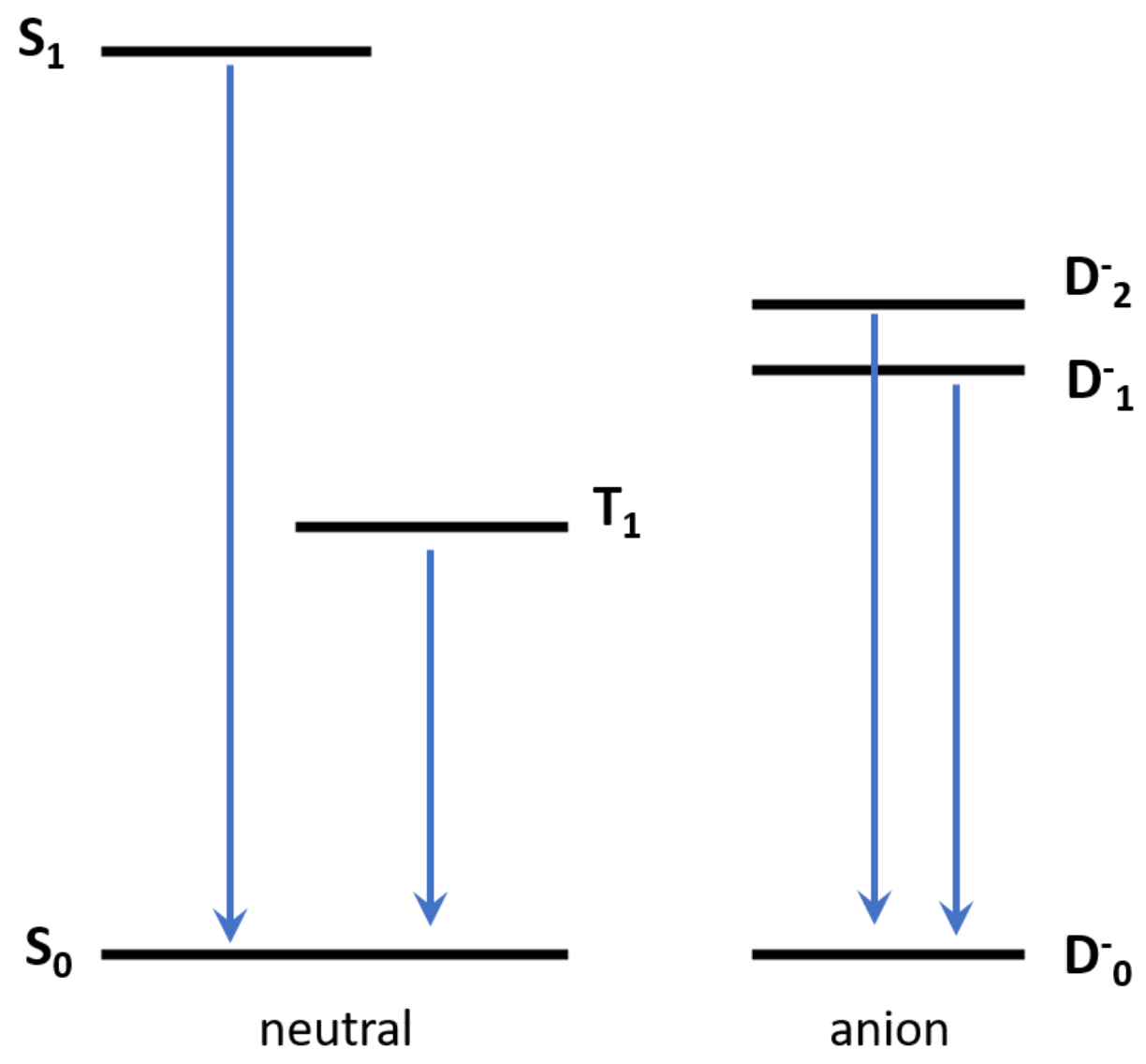

**Figure S7**. Energy level scheme of the PTCDA monomer excited states of neutral and anion (see Table S1).

| | **charge (S)** | **Relative Energy** | **charge localization** | | | | **Scheme** |
|---|---|---|---|---|---|---|---|
| | **e$^-$** | **eV** | **mol.a** | **mol.b** | **mol.c** | **mol.d** | |
| **monomer** | -1 (1/2) | -2.65 | - | - | - | - | 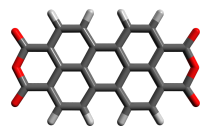 |
| **dimer parallel** | -2 (1 or 0) | -3.63 | -1.00 | -1.00 | - | - | 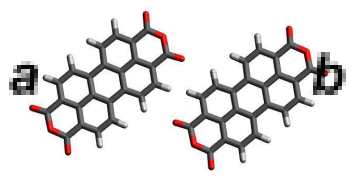 |
| **dimer perpen-dicular** | -2 (1 or 0) | -3.64 | -0.97 | -1.03 | - | - | 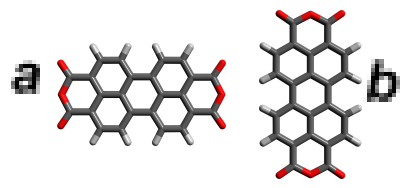 |
| **trimer** | -2 (1 or 0) | -4.67 | -0.97 | -0.06 | -0.97 | - | 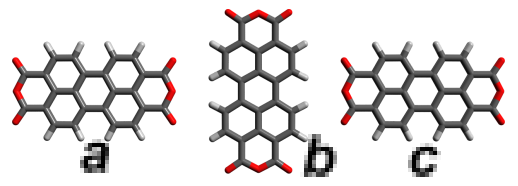 |
| | -3 (3/2 or 1/2) | -4.09 | -0.97 | -1.06 | -0.97 | - | |
| **tetramer** | -3 (3/2) | -3.95 | -0.75 | -0.75 | -0.75 | -0.75 | 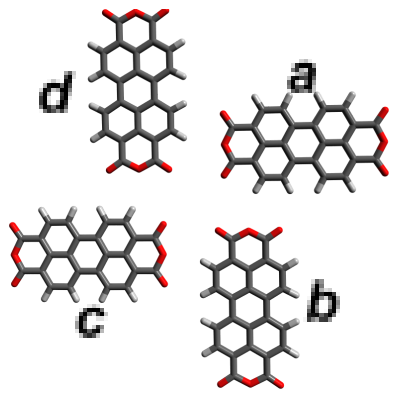 |
| | -3 (1/2) | -3.98 | -0.50 | -1.00 | -0.50 | -1.00 | |
| | -2 (1 or 0) | -4.01 | -0.01 | -0.99 | -0.01 | -0.99 | |

**Table S2.** Total charge and spin of each system, absolute energy difference with respect to the neutral ground state, amount of charge localized on each molecule of the aggregate (labeled according to the scheme in the last column), computed at the wB97XD/6-31G* level of theory.

| dimer parallel S = 1 | E/eV (*f*) | wf.[a] |
|---|---|---|
| $T_1^{2-}$ (Bu) | 1.852 (0.296) | 0.39 $L^{+}_{\alpha} \rightarrow L{+}1^{+}_{\alpha}$ |
| | | 0.35 $L^{-}_{\alpha} \rightarrow L{+}1^{-}_{\alpha}$ |
| | | 0.11 $H^{-}_{\beta} \rightarrow L^{-}_{\beta}$ |
| | | 0.11 $H^{+}_{\beta} \rightarrow L^{+}_{\beta}$ |
| | | 0.01 $L^{-}_{\alpha} \rightarrow L{+}2^{-}_{\alpha}$ |
| $T_2^{2-}$ (Ag) | 1.857 (0) | 0.36 $L^{+}_{\alpha} \rightarrow L{+}1^{-}_{\alpha}$ |
| | | 0.34 $L^{-}_{\alpha} \rightarrow L{+}1^{+}_{\alpha}$ |
| | | 0.11 $H^{+}_{\beta} \rightarrow L^{-}_{\beta}$ |
| | | 0.11 $H^{-}_{\beta} \rightarrow L^{+}_{\beta}$ |
| | | 0.03 $L^{-}_{\alpha} \rightarrow L{+}2^{+}_{\alpha}$ |
| | | 0.01 $L^{+}_{\alpha} \rightarrow L{+}2^{-}_{\alpha}$ |
| $T_3^{2-}$ (Bu) | 1.906 (0.165) | 0.44 $L^{-}_{\alpha} \rightarrow L{+}2^{-}_{\alpha}$ |
| | | 0.47 $L^{+}_{\alpha} \rightarrow L{+}2^{+}_{\alpha}$ |
| | | 0.01 $L^{-}_{\alpha} \rightarrow L{+}1^{-}_{\alpha}$ |
| $T_4^{2-}$ (Ag) | 1.923 (0) | 0.43 $L^{-}_{\alpha} \rightarrow L{+}2^{+}_{\alpha}$ |
| | | 0.43 $L^{+}_{\alpha} \rightarrow L{+}2^{-}_{\alpha}$ |
| | | 0.01 $L^{-}_{\alpha} \rightarrow L{+}1^{+}_{\alpha}$ |

[a] For the orbitals notation, refer to Fig. S8.

**Table S3.** Vertical excited states of the parallel dimer (total charge -2, triplet spin state): excitation energies (E), oscillator strengths (*f*) and wavefunction composition (wf) including coefficients and aggregate orbitals involved in the dominant excitations (coefficients ≥ 0.01).

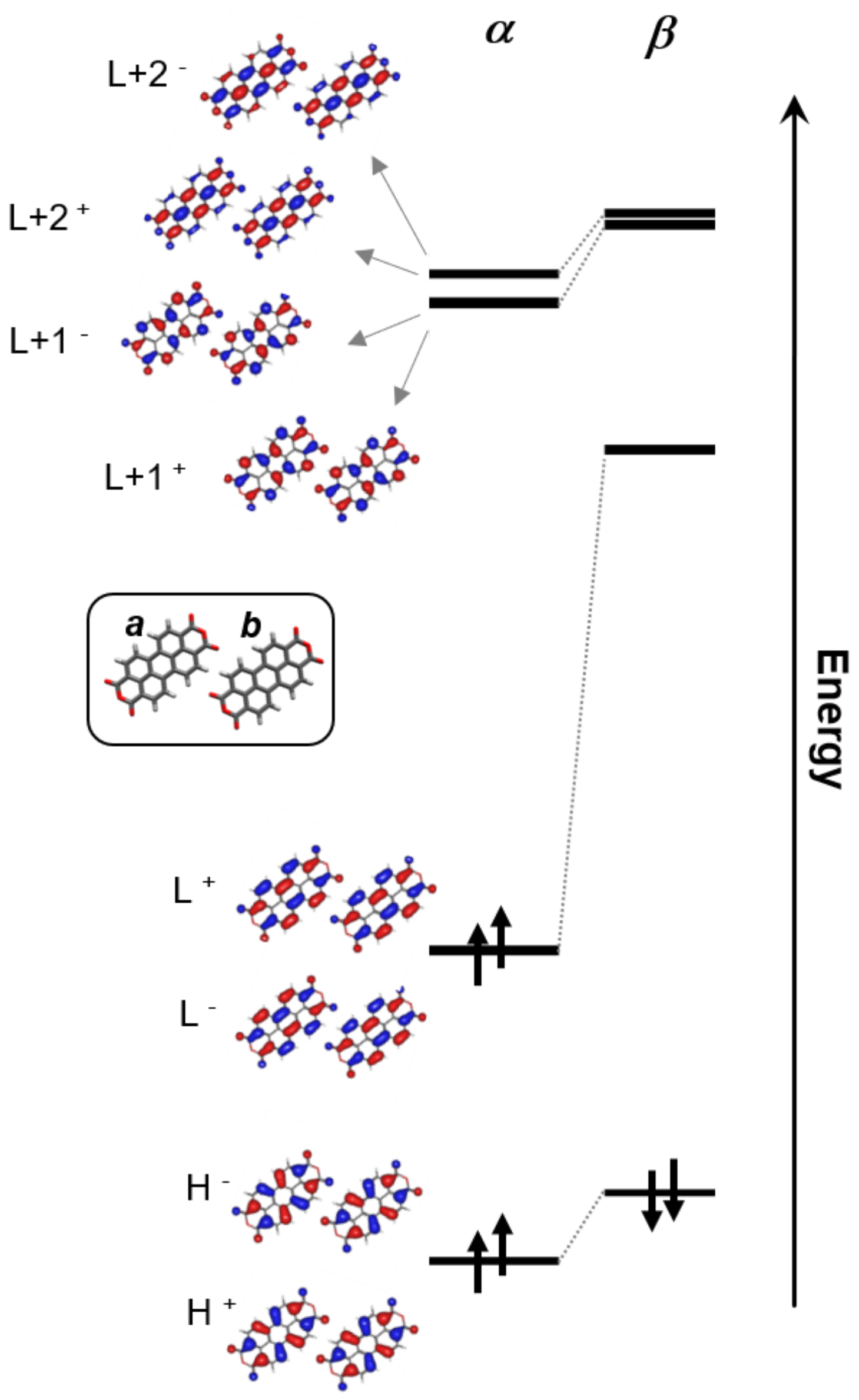

**Figure S8.** Scheme of the ground state aggregate orbitals involved in the relevant vertical excited states of the parallel dimer (total charge -2, triplet spin state).

| dimer perpendicular S = 1 | E/eV (*f*) | wf.[a] |
|---|---|---|
| $T_1^{2-}$ (B2) | 1.852 (0.108) | 0.73 $L^{a}_{\alpha} \rightarrow L{+}1^{a}_{\alpha}$ |
| | | 0.22 $H^{a}_{\beta} \rightarrow L^{a}_{\beta}$ |
| | | 0.01 $L^{b}_{\alpha} \rightarrow L{+}2^{b}_{\alpha}$ |
| $T_2^{2-}$ (A1) | 1.872 (0.189) | 0.70 $L^{b}_{\alpha} \rightarrow L{+}1^{b}_{\alpha}$ |
| | | 0.22 $H^{b}_{\beta} \rightarrow L^{b}_{\beta}$ |
| | | 0.05 $L^{a}_{\alpha} \rightarrow L{+}2^{a}_{\alpha}$ |
| $T_3^{2-}$ (A1) | 1.915 (0.052) | 0.88 $L^{a}_{\alpha} \rightarrow L{+}2^{a}_{\alpha}$ |
| | | 0.03 $L^{b}_{\alpha} \rightarrow L{+}1^{b}_{\alpha}$ |
| | | 0.02 $L^{b}_{\beta} \rightarrow L{+}2^{b}_{\beta}$ |
| $T_4^{2-}$ (B2) | 1.920 (0.081) | 0.91 $L^{b}_{\alpha} \rightarrow L{+}2^{b}_{\alpha}$ |

[a] For the orbitals notation, refer to Fig.S9.

**Table S4.** Vertical excited states of the perpendicular dimer (total charge -2, triplet spin state): excitation energies (E), oscillator strengths (*f*) and wavefunction composition (wf) including coefficients and aggregate orbitals involved in the dominant excitations (coefficients ≥ 0.01).

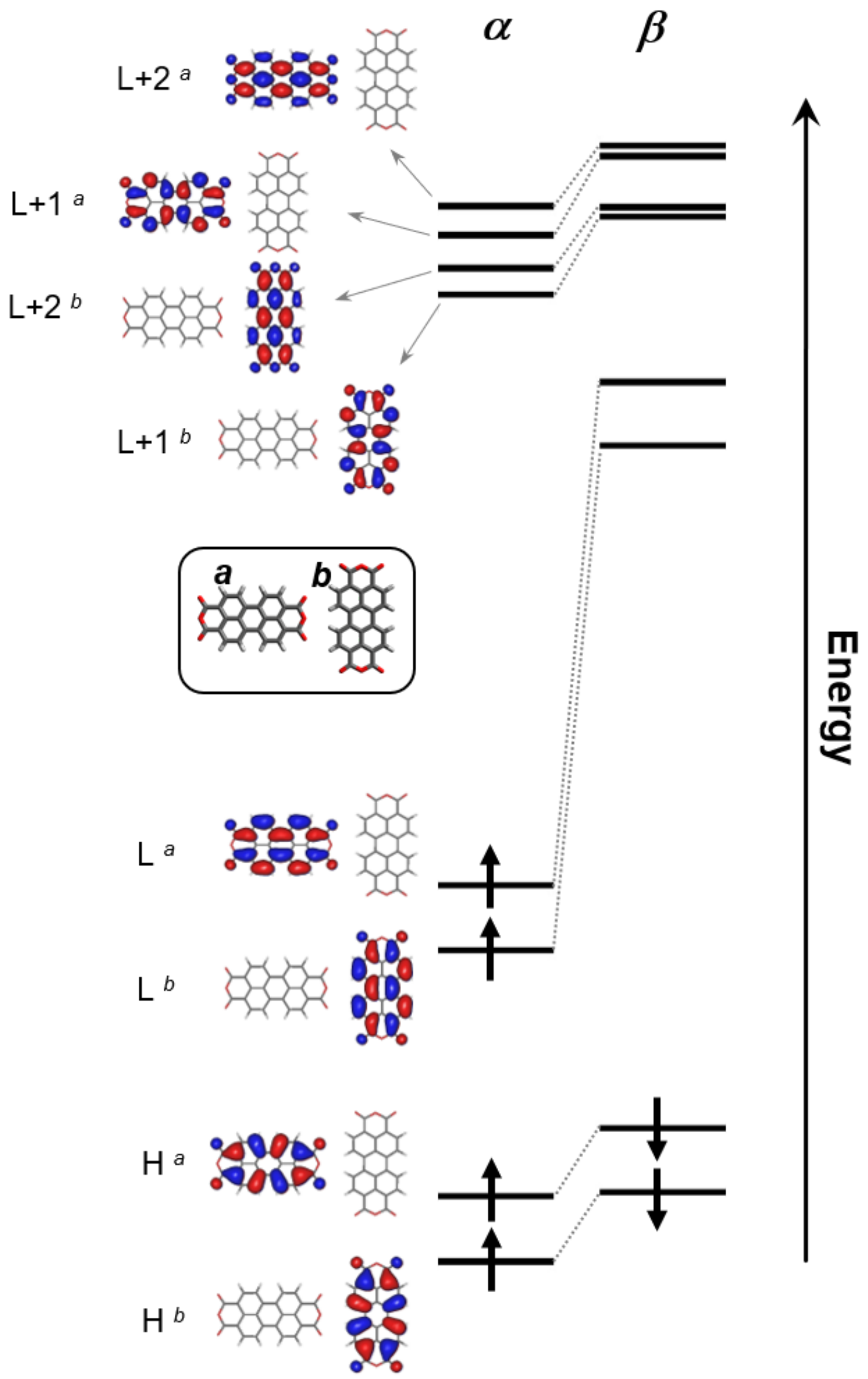

**Figure S9.** Scheme of the ground state aggregate orbitals involved in the relevant vertical excited states of the perpendicular dimer (total charge -2, triplet spin state).

| trimer [a] S = 1 | E/eV (*f*) | wf.[b] |
|---|---|---|
| $T_1^{2-}$ (B1u) | 1.863 (0.309) | 0.37 $L^-_\alpha \rightarrow L+1^-_\alpha$ |
| | | 0.37 $L^+_\alpha \rightarrow L+1^+_\alpha$ |
| | | 0.11 $H^+_\beta \rightarrow L^+_\beta$ |
| | | 0.11 $H^-_\beta \rightarrow L^-_\beta$ |
| $T_2^{2-}$ (Ag) | 1.878 (0) | 0.36 $L^-_\alpha \rightarrow L+1^+_\alpha$ |
| | | 0.36 $L^+_\alpha \rightarrow L+1^-_\alpha$ |
| | | 0.12 $H^+_\beta \rightarrow L^-_\beta$ |
| | | 0.12 $H^-_\beta \rightarrow L^+_\beta$ |
| $T_3^{2-}$ (B3g) | 1.916 (0) | 0.46 $L^-_\alpha \rightarrow L+2^+_\alpha$ |
| | | 0.46 $L^+_\alpha \rightarrow L+2^-_\alpha$ |
| $T_4^{2-}$ (B2u) | 1.917 (0.130) | 0.46 $L^-_\alpha \rightarrow L+2^-_\alpha$ |
| | | 0.46 $L^+_\alpha \rightarrow L+2^+_\alpha$ |

[a] The excited state ordering from calculations is 4-7 (the lower states were discarded).

[b] For the orbitals notation, refer to Fig. S10.

**Table S5.** Vertical excited states of the trimer (total charge -2, triplet spin state): excitation energies (E), oscillator strengths (*f*) and wavefunction composition (wf) including coefficients and aggregate orbitals involved in the dominant excitations (coefficients ≥ 0.01).

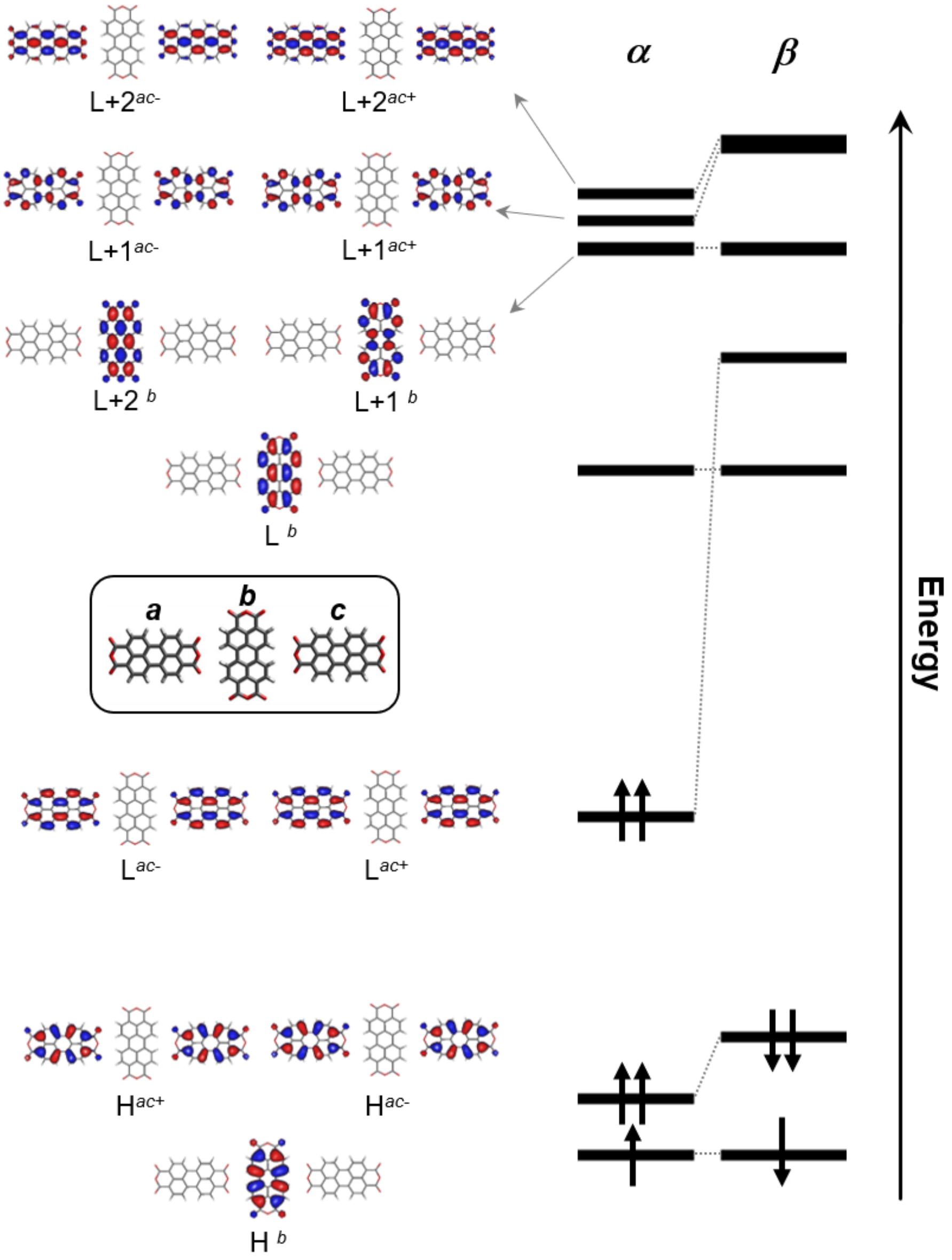

**Figure S10.** Scheme of the ground state aggregate orbitals involved in the relevant vertical excited states of the trimer (total charge -2, triplet spin state).

| trimer S = 3/2 | E/eV (f) | wf. [a] |
|---|---|---|
| $Q_1^{3-}$ (B2u) | 1.841(0.094) | 0.74 $L^{b}_{\alpha} \rightarrow L{+}1^{b}_{\alpha}$ |
| | | 0.21 $H^{b}_{\beta} \rightarrow L^{b}_{\beta}$ |
| $Q_2^{3-}$ (B1u) | 1.869(0.413) | 0.35 $L^{b}_{\alpha} \rightarrow L{+}2^{b}_{\alpha}$ |
| | | 0.24 $L^{ac-}_{\alpha} \rightarrow L{+}1^{ac-}_{\alpha}$ |
| | | 0.24 $L^{ac+}_{\alpha} \rightarrow L{+}1^{ac+}_{\alpha}$ |
| | | 0.06 $H^{ac+}_{\beta} \rightarrow L^{ac+}_{\beta}$ |
| | | 0.06 $H^{ac-}_{\beta} \rightarrow L^{ac-}_{\beta}$ |
| $Q_3^{3-}$ (Ag) | 1.884(0) | 0.36 $L^{ac-}_{\alpha} \rightarrow L{+}1^{ac+}_{\alpha}$ |
| | | 0.36 $L^{ac+}_{\alpha} \rightarrow L{+}1^{ac-}_{\alpha}$ |
| | | 0.12 $H^{ac+}_{\beta} \rightarrow L^{ac-}_{\beta}$ |
| | | 0.12 $H^{ac-}_{\beta} \rightarrow L^{ac+}_{\beta}$ |
| $Q_4^{3-}$ (B1u) | 1.897(0.001) | 0.58 $L^{b}_{\alpha} \rightarrow L{+}2^{b}_{\alpha}$ |
| | | 0.13 $L^{ac-}_{\alpha} \rightarrow L{+}1^{ac-}_{\alpha}$ |
| | | 0.13 $L^{ac+}_{\alpha} \rightarrow L{+}1^{ac+}_{\alpha}$ |
| | | 0.06 $H^{ac+}_{\beta} \rightarrow L^{ac+}_{\beta}$ |
| | | 0.06 $H^{ac-}_{\beta} \rightarrow L^{ac-}_{\beta}$ |
| $Q_5^{3-}$ (B3g) | 1.917(0) | 0.46 $L^{ac-}_{\alpha} \rightarrow L{+}2^{ac+}_{\alpha}$ |
| | | 0.46 $L^{ac+}_{\alpha} \rightarrow L{+}2^{ac-}_{\alpha}$ |
| $Q_6^{3-}$ (B2u) | 1.919(0.157) | 0.45 $L^{ac-}_{\alpha} \rightarrow L{+}2^{ac-}_{\alpha}$ |
| | | 0.45 $L^{ac+}_{\alpha} \rightarrow L{+}2^{ac+}_{\alpha}$ |
| | | 0.01 $L^{b}_{\alpha} \rightarrow L{+}1^{b}_{\alpha}$ |

[a] for the orbitals notation, refer to Fig. S11.

**Table S6.** Vertical excited states of the trimer (total charge -3, quartet spin state): excitation energies (E), oscillator strengths (*f*) and wavefunction composition (wf) including coefficients and aggregate orbitals involved in the dominant excitations (coefficients ≥ 0.01).

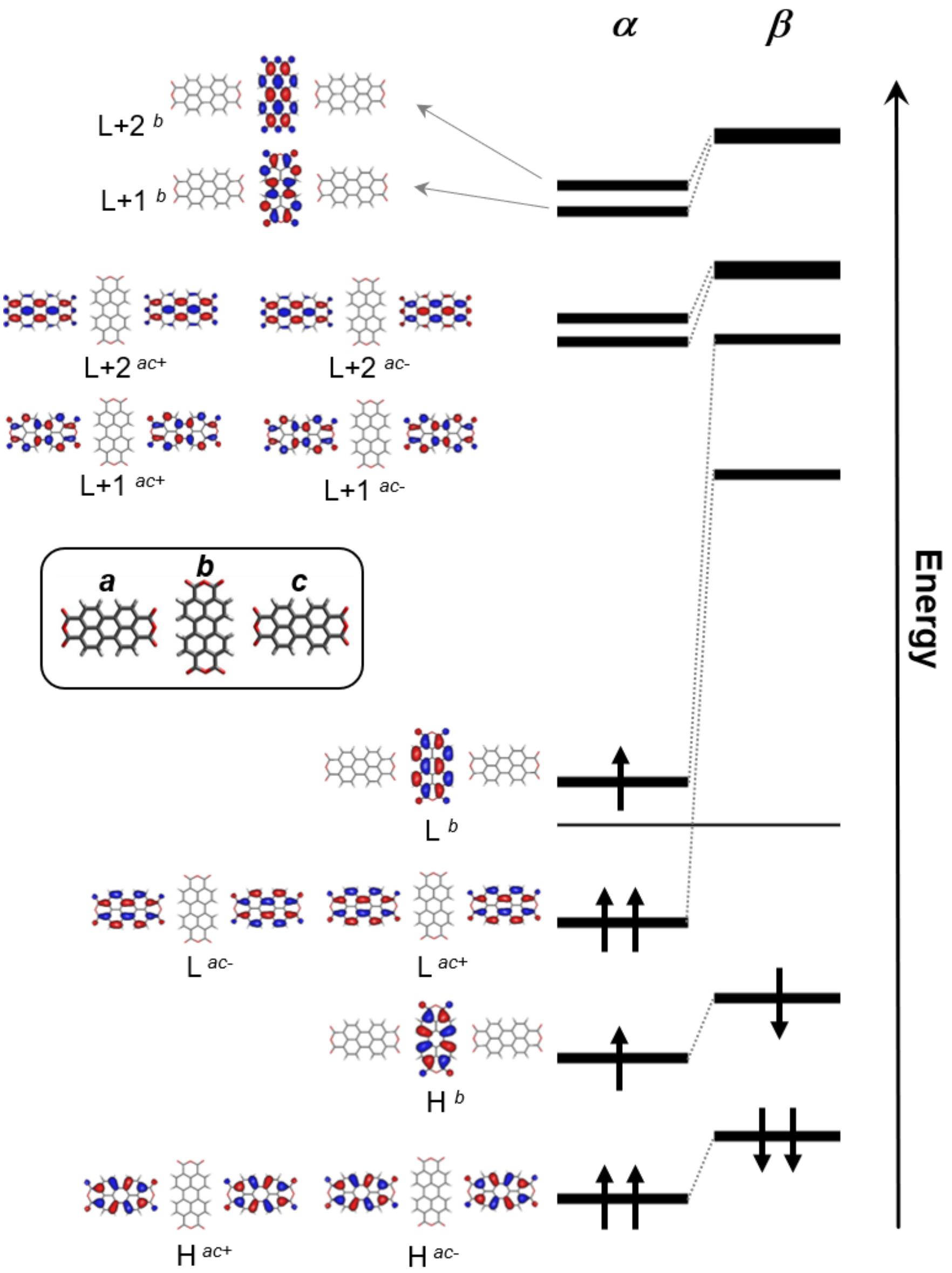

**Figure S11.** Scheme of the ground state aggregate orbitals involved in the relevant vertical excited states of the trimer (total charge -3, quartet spin state).

| **Tetramer** [a] **S = 3/2** | **E/eV (f)** | **wf.** [b] |
|---|---|---|
| $Q_1^{3-}$ (Bg) | 1.824(0) | 0.20 $L^{t-}_{\alpha} \rightarrow L^{t+}_{\alpha}$ |
| | | 0.20 $L^{ac-}_{\alpha} \rightarrow L{+}1^{ac-}_{\alpha}$ |
| | | 0.20 $L^{bd-}_{\alpha} \rightarrow L{+}1^{bd-}_{\alpha}$ |
| | | 0.08 $H^{t-}_{\alpha} \rightarrow L^{t+}_{\alpha}$ |
| | | 0.07 $H^{t+}_{\beta} \rightarrow L^{t-}_{\beta}$ |
| | | 0.07 $H^{t-}_{\beta} \rightarrow L^{t+}_{\beta}$ |
| | | 0.06 $H^{ac-}_{\beta} \rightarrow L^{ac-}_{\beta}$ |
| | | 0.06 $H^{bd-}_{\beta} \rightarrow L^{bd-}_{\beta}$ |
| $Q_2^{3-}$ (Eu) | 1.826(0.185) | 0.17 $L^{bd-}_{\alpha} \rightarrow L{+}1^{t+}_{\alpha}$ |
| | | 0.16 $L^{t-}_{\alpha} \rightarrow L{+}1^{bd-}_{\alpha}$ |
| | | 0.15 $L^{bd-}_{\alpha} \rightarrow L{+}1^{t-}_{\alpha}$ |
| | | 0.06 $H^{bd-}_{\beta} \rightarrow L^{t-}_{\beta}$ |
| | | 0.06 $H^{ac-}_{\alpha} \rightarrow L^{t+}_{\alpha}$ |
| | | 0.06 $H^{t-}_{\beta} \rightarrow L^{bd+}_{\beta}$ |
| | | 0.05 $H^{t+}_{\beta} \rightarrow L^{bd-}_{\beta}$ |
| | | 0.05 $L^{ac-}_{\alpha} \rightarrow L{+}1^{t-}_{\alpha}$ |
| $Q_3^{3-}$ (Eu) | 1.826(0.185) | 0.17 $L^{ac-}_{\alpha} \rightarrow L{+}1^{t+}_{\alpha}$ |
| | | 0.16 $L^{t-}_{\alpha} \rightarrow L{+}1^{ac+}_{\alpha}$ |
| | | 0.15 $L^{ac-}_{\alpha} \rightarrow L{+}1^{t-}_{\alpha}$ |
| | | 0.06 $H^{ac-}_{\beta} \rightarrow L^{t-}_{\beta}$ |
| | | 0.06 $H^{bd-}_{\alpha} \rightarrow L^{t+}_{\alpha}$ |
| | | 0.06 $H^{t-}_{\beta} \rightarrow L^{ac-}_{\beta}$ |
| | | 0.05 $H^{t+}_{\beta} \rightarrow L^{ac-}_{\beta}$ |
| | | 0.05 $L^{bd-}_{\alpha} \rightarrow L{+}1^{t-}_{\alpha}$ |
| $Q_4^{3-}$ | 1.828(0) | 0.19 $L^{ac-}_{\alpha} \rightarrow L{+}1^{ac+}_{\alpha}$ |
| | | 0.19 $L^{bd-}_{\alpha} \rightarrow L{+}1^{bd-}_{\alpha}$ |
| | | 0.19 $L^{t-}_{\alpha} \rightarrow L{+}1^{t-}_{\alpha}$ |
| | | 0.07 $H^{t+}_{\alpha} \rightarrow L^{t+}_{\alpha}$ |
| | | 0.07 $H^{t-}_{\beta} \rightarrow L^{t-}_{\beta}$ |
| | | 0.07 $H^{t+}_{\beta} \rightarrow L^{t+}_{\beta}$ |
| | | 0.07 $H^{ac-}_{\beta} \rightarrow L^{ac-}_{\beta}$ |
| | | 0.07 $H^{bd-}_{\beta} \rightarrow L^{bd-}_{\beta}$ |

[a] The excited state ordering from calculations is 4-7 (lower states were discarded).

[b] For the orbitals notation refer to Fig.S12. The notation “$ac^{+}bd^{+}$” has been shortened to “t”.

**Table S7.** Vertical excited states of the tetramer (total charge -3, quartet spin state): excitation energies (E), oscillator strengths (*f*) and wavefunction composition (wf) including coefficients and aggregate orbitals involved in the dominant excitations (coefficients ≥ 0.05).

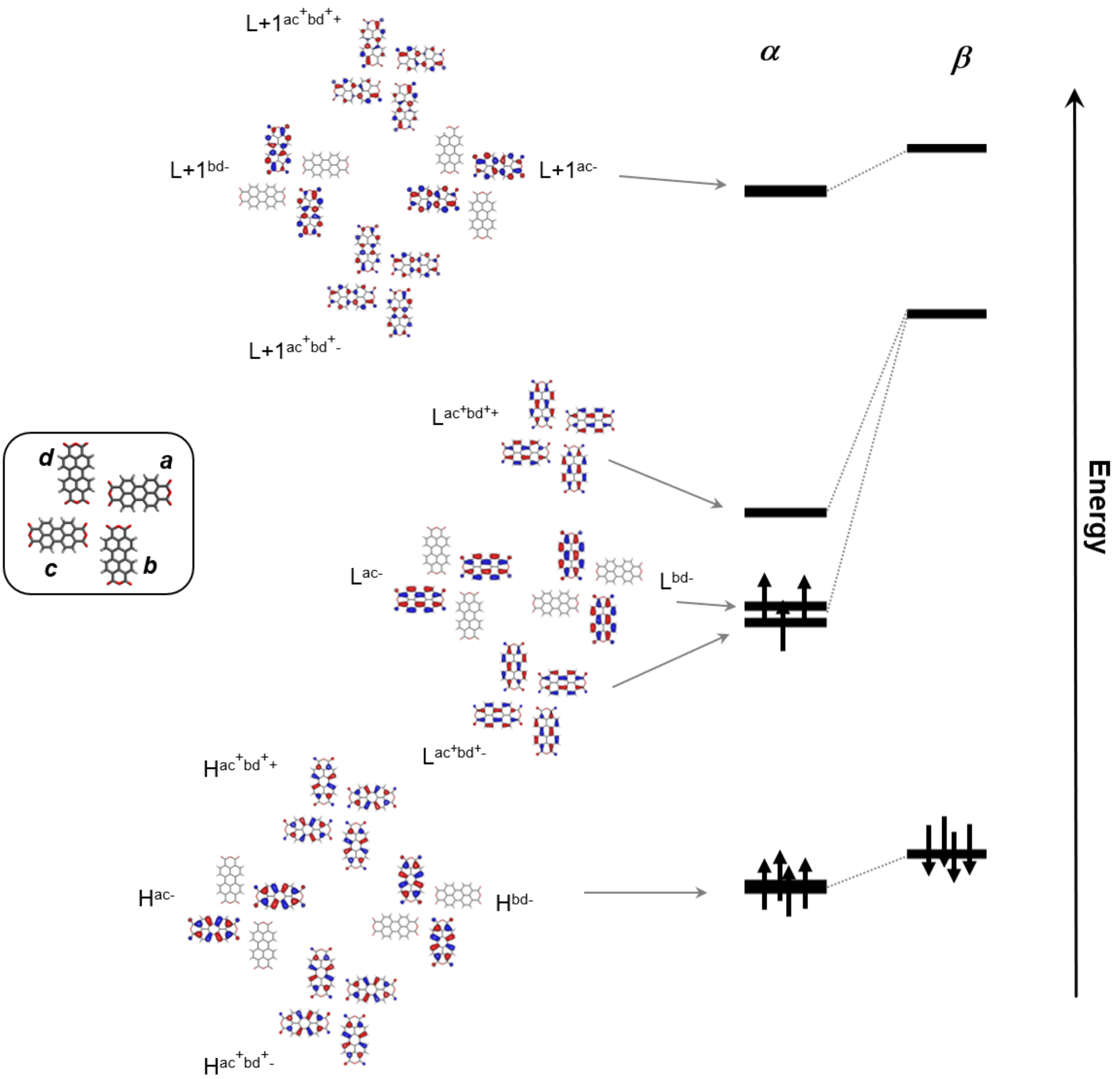

**Figure S12.** Scheme of the ground state aggregate orbitals involved in the relevant vertical excited states of the tetramer (total charge -3, quartet spin state).

| Tetramer [a] S = 1 | E/eV (*f*) | wf.[b] |
|---|---|---|
| $T_1^{2-}$ (Bg) | 1.845 (0.283) | 0.37 $L_\alpha^{ac-} \rightarrow L+1_\alpha^{ac-}$<br>0.37 $L_\alpha^{ac+} \rightarrow L+1_\alpha^{ac+}$<br>0.09 $H_\beta^{ac+} \rightarrow L_\beta^{ac+}$<br>0.09 $H_\beta^{ac-} \rightarrow L_\beta^{ac-}$<br>0.02 $L_\alpha^{ac-} \rightarrow L+2_\alpha^{ac-}$<br>0.02 $L_\alpha^{ac+} \rightarrow L+2_\alpha^{ac+}$ |

[a] The excited state ordering number from calculations is 7 (the lower states were discarded).

[b] For the orbitals notation, refer to Fig.S13.

**Table S8.** Vertical excited states of the tetramer (total charge -2, triplet spin state): excitation energies (E), oscillator strengths (*f*) and wavefunction composition (wf) including coefficients and aggregate orbitals involved in the dominant excitations (coefficients ≥ 0.05).

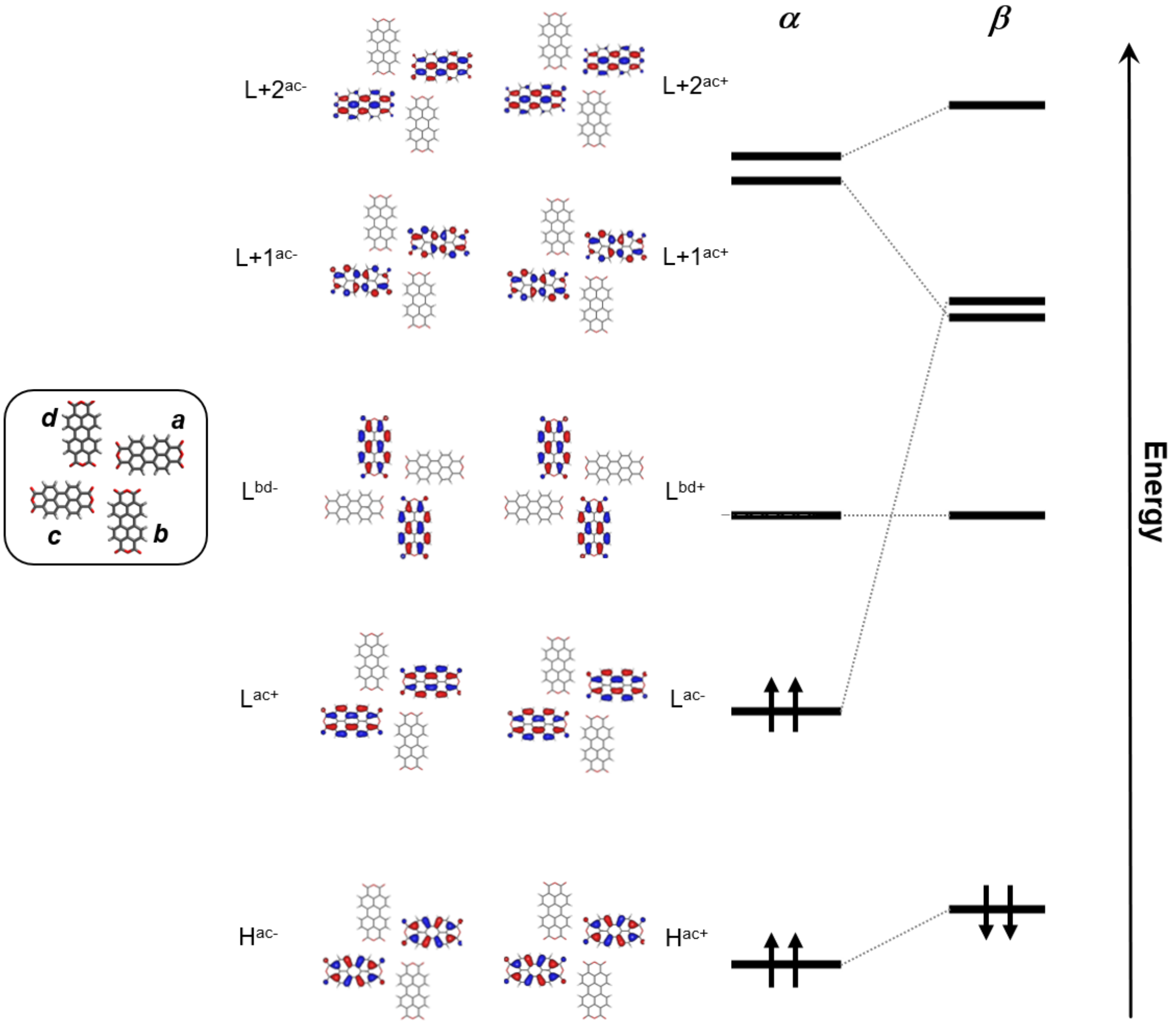

**Figure S13.** Scheme of the ground state aggregate orbitals involved in the relevant vertical excited states of the tetramer (total charge -2, triplet spin state).

**Supporting Information references**